# Machine Learning-Driven Materials Discovery: Unlocking Next-Generation Functional Materials – A review

Dilshod Nematov[a,b *] and Mirabbos Hojamberdiev[c,*]

*[a] School of Optoelectronic Engineering & CQUPT-BUL Innovation Institute, Chongqing University of Posts and Telecommunications, Chongqing 400065, China*

*[b] S.U. Umarov Physical-Technical Institute of National Academy of Sciences of Tajikistan, 299/1 Ayni Street, 734042 Dushanbe, Tajikistan*

*[c] Mads Clausen Institute, University of Southern Denmark, Alison 2, 6400 Sønderborg, Denmark*

*Corresponding authors: dilnem@mail.ru (D. Nematov) and mirabbos@mci.sdu.dk (M. Hojamberdiev)

**ABSTRACT:**

The rapid advancement of machine learning and artificial intelligence (AI)-driven techniques is revolutionizing materials discovery, property prediction, and material design by minimizing human intervention and accelerating scientific progress. This review provides a comprehensive overview of smart, machine learning (ML)-driven approaches, emphasizing their role in predicting material properties, discovering novel compounds, and optimizing material structures. Key methodologies in this field include deep learning, graph neural networks, Bayesian optimization, and automated generative models (GANs, VAEs). These approaches enable the autonomous design of materials with tailored functionalities. By leveraging AutoML frameworks (AutoGluon, TPOT, and H2O.ai), researchers can automate the model selection, hyperparameter tuning, and feature engineering, significantly improving the efficiency of materials informatics. Furthermore, the integration of AI-driven robotic laboratories and high-throughput computing has established a fully automated pipeline for rapid synthesis and experimental validation, drastically reducing the time and cost of material discovery. This review highlights real-world applications of automated ML-driven approaches in predicting mechanical, thermal, electrical, and optical properties of materials, demonstrating successful cases in superconductors, catalysts, photovoltaics, and energy storage systems. We also address key challenges, such as data quality, interpretability, and the integration of AutoML with quantum computing, which are essential for future advancements. Ultimately, combining AI with automated experimentation and computational modeling is transforming the way materials are discovered and optimized. This synergy paves the way for new innovations in energy, electronics, and nanotechnology.

## 1. Introduction

Machine learning (ML) has become a transformative tool in modern materials science, offering new opportunities to predict material properties, design novel compounds, and optimize performance. Traditional empirical experiments and classical theoretical modeling are time-consuming and costly [1-3]. With the rapid growth of data from experiments, simulations, and databases (Materials Project, OQMD, AFLOW, NOMAD), conventional methods struggle to meet current research demands. ML overcomes these challenges by analyzing large datasets and revealing complex relationships between chemical composition, microstructural features, and material properties [4,5]. A major limitation of traditional methods is scalability. While density functional theory (DFT) and molecular dynamics (MD) simulations deliver high accuracy, they are computationally intensive and slow, especially for complex multicomponent systems [6,7]. Moreover, the vast chemical space makes experimental testing of every candidate impractical, hindering innovation [3,8,9]. ML addresses these issues by training models on extensive datasets to automate the property prediction and reduce experimental efforts [10]. It also integrates diverse data sources, as modern databases provide a robust foundation for neural network training [11]. Deep learning techniques, such as convolutional neural networks (CNNs) and graph neural networks (GNNs), have achieved highly accurate predictions even for complex crystalline structures [12,13], thereby shifting material design from lengthy experimental cycles to targeted creation with predefined properties [14]. Before the advent of data-driven methods, material discovery relied heavily on trial-and-error experimental techniques and first-principles calculations, such as density functional theory (DFT), molecular dynamics (MD), and quantum chemistry models. These approaches, while accurate, are limited by time-consuming simulations and high computational cost, especially when exploring large compositional or structural design spaces [1-9]. Traditional combinatorial synthesis and high-throughput screening have also been widely used, but they face limitations in scalability and generalizability.

Modern ML methods utilize extensive training datasets from large-scale materials databases to develop accurate predictive models. However, choosing an optimal model remains challenging, underfitting fails to capture complex relationships while overfitting yields overly specialized models [15,16]. A key advantage of ML is its ability to integrate diverse data sources, enhancing our overall understanding of material properties. By combining high-throughput simulations, experimental measurements, and database information, researchers develop robust models that predict material characteristics under varied conditions [17,18]. For example, deep learning has been effectively used to predict thermoelectric properties, which is crucial for next-generation energy generators [19]. ML also offers cost efficiency. Traditional methods like DFT demand significant computational resources, limiting large-scale screening. ML models, trained on existing data, provide rapid preliminary assessments so that only promising candidates undergo more detailed analysis [20,21]. Furthermore, ML spurs hypothesis generation and innovation. Modern algorithms, such as generative adversarial networks (GANs) and variational autoencoders (VAEs), can generate novel chemical compositions meeting specific criteria, leading to the discovery of previously unexplored material classes [22,23]. Recent surveys summarize how GANs, VAEs, and diffusion models are being adapted for materials design, detailing data representations, evaluation protocols, and emerging best practices in inverse design. In particular, equivariant diffusion and transformer-based generators have demonstrated the ability to propose DFT-relaxable inorganic structures and recover known materials distributions, providing a principled route to generate candidates prior to targeted validation [4, 21, 23-25]. These methods are critical for developing functional materials for emerging technologies, including quantum computing, energy-efficient batteries, and advanced photocatalysts [24].

Recent studies show that integrating ML with traditional computational and experimental methods produces hybrid models with enhanced prediction accuracy. Such models are already applied in designing semiconductors, batteries, solar cells, and catalysts [24–26]. For example, deep learning combined with DFT data has improved solar cell efficiency, advancing renewable energy technologies [27,28]. Modern computational resources like GPUs and TPUs accelerate neural network training, enabling the development of complex models and paving the way for “smart” laboratories where ML-driven systems conduct real-time material synthesis and optimization [29,30]. Moreover, ML promotes interdisciplinary collaboration among materials scientists, computational mathematicians, and informatics experts, leading to novel hybrid models with superior predictive performance [31,32].

Integrating ML in materials science opens new avenues for designing and discovering functional materials. Its benefits from the reduced time and cost, high prediction accuracy, universal model applicability, and effective integration of diverse data surpass those of traditional methods. Advancements in computational capabilities, improved algorithms, and expanding materials databases indicate that ML will become an indispensable part of materials research, ultimately leading to efficient, sustainable materials for energy, electronics, medicine, and beyond [33-37]. This review aims to systematize modern ML approaches in materials science, examine existing challenges, and highlight the advantages of ML in predicting, designing, and discovering novel functional materials. The interdisciplinary nature of this field and rapid technological progress promise groundbreaking advancements, making it highly relevant to both the scientific community and industry [34–39]. ML-driven discovery is increasingly being combined with laboratory experimentation in circular economy frameworks. For instance, value-added materials derived from industrial waste streams are optimized using ML models to identify secondary raw materials with functional properties [34-45]. Integration with robotic labs allows bridging of lab-to-industry transition for water purification membranes, $CO_2$ capture materials, and catalysts for clean energy.

## 2. Key ML algorithms for materials science

The re-emergence of ML is driven by increased data availability, computational advances, and enhanced computing power [37–40]. Initially rooted in statistical learning, ML now permeates physics, chemistry, and materials science. It uses historical data (inputs) to generate predictions (outputs) via various algorithms. Performance depends on dataset size and computational efficiency [40]. Unlike traditional experimental or simulation-based approaches, ML focuses on data processing and statistical analysis while integrating ideas from computer science, statistics, and optimization. Standard ML tasks include classification, regression, ranking, clustering, and dimensionality reduction [40]. Machine learning has become a powerful tool in materials science, enabling the prediction, optimization, and discovery of new materials. In modern materials research, ML algorithms are applied across a wide range of fields, from developing new materials for solar cells and batteries to modeling microstructural characteristics of polymers and alloys. The primary algorithms used in this field can be categorized based on their characteristics, advantages, and application areas. Table 1 summarizes the main machine learning algorithms, their applicability, and key features. Notably, many algorithms, such as random forests, can be used for both classification and regression.

**Table 1.** Application areas and characteristics of key machine learning algorithms.

| **Method** | **Category** | **Applicable scenarios and functions** |
|---|---|---|
| Support Vector Regression (SVR) | Regression | SVR is a nonlinear algorithm that works well with small datasets and is resistant to overfitting. |
| Artificial Neural Network (ANN) | Regression | Requires large datasets, has self-learning capabilities, and is robust to failures, but interpretability is weak. |
| Linear Regression | Regression | Requires strict assumptions and linearly correlated data; offers fast modeling and good interpretability. |
| Logistic Regression | Regression | Widely used for classification tasks but cannot handle multiple feature-variable relationships effectively. |
| Kernel Ridge Regression (KRR) | Regression | Handles nonlinear data but has slower prediction speed compared to SVR with large datasets. |
| Support Vector Classification (SVC) | Classification | SVC, also known as a maximum-margin classifier, is particularly effective for binary classification tasks. |
| K-Nearest Neighbors (KNN) | Classification | Suitable for multiclass classification but computationally expensive with high sample balance requirements. |
| Decision Tree (DT) | Classification | Handles missing attributes well and offers good interpretability but lacks online learning support and is prone to overfitting. |
| Random Forest (RF) | Classification | Inherits DT advantages while preventing overfitting, even with small noise levels. |
| K-Means Clustering | Clustering | A classical clustering algorithm known for its simplicity and speed but sensitive to initial conditions. |
| Hierarchical Cluster Analysis (HCA) | Clustering | Constructs cluster hierarchies in a single process but is computationally intensive. |
| Hidden Markov Model (HMM) | Clustering | A key stochastic model for signal processing with broad applications in pattern recognition. |
| Gaussian Process Regression (GPR) | Regression | Bayesian, calibrated uncertainty |
| Gradient Boosting (XGBoost/CatBoost) | Ensemble (Reg./Clf.) | Strong accuracy; needs tuning |
| Convolutional Neural Networks (CNN) | Deep Learning | Images/spectra/microstructures |
| Recurrent Neural Networks (RNN/LSTM) | Deep Learning | Equences/time series (degradation) |
| Graph Neural Networks (GNN) | Deep Learning | Graphs of atoms/bonds; structure-aware |
| Autoencoders (AE) | Unsupervised | Dimensionality reduction/denoising; latent features |
| PCA / t-SNE / UMAP | Unsupervised (DR) | Visualize/condense high-dim data |
| DBSCAN / OPTICS | Clustering | Density-based; outlier detection; no k needed |
| Generative Models (VAE, GAN) | Deep Learning | Inverse design; candidate generation. |

Regression methods (linear regression, SVR, Gaussian Processes, gradient boosting) are extensively used in materials science to predict continuous mechanical, thermal, electronic, and optical properties [41–44]. Linear regression, despite its simplicity, effectively predicts Young's modulus [3], while SVR handles nonlinear dependencies by using kernel functions [42]. Gaussian Processes incorporate uncertainty estimates [43], and gradient boosting combines weak learners to improve accuracy [44]. Classification methods are crucial when categorizing materials by properties, structure, or composition. SVM efficiently separates data into classes [5,45], Random Forest leverages ensembles of decision trees while assessing feature importance [46], and neural networks are versatile for tasks like classifying crystalline structures [47]. Ensemble methods boost prediction accuracy by combining multiple models. Gradient boosting excels in tasks such as predicting molecular energy levels [28, 48]; its popular implementations include XGBoost for thermoelectric properties [49] and CatBoost for categorical data [50]. Deep learning is rapidly evolving. CNNs are widely applied for microstructure image analysis [2,51], RNNs handle time-series data like material degradation [52], and GANs can generate novel crystalline structures, expanding the search space for new compounds [53]. Clustering methods uncover hidden patterns in large datasets. K-means groups materials with similar mechanical properties [54], while hierarchical clustering visualizes data in dendrograms [55]. Algorithms like DBSCAN and OPTICS identify clusters of arbitrary shapes, aiding studies of amorphous materials. Dimensionality reduction simplifies high-dimensional data. PCA extracts principal components for efficient analysis [56], while t-SNE and UMAP handle nonlinear relationships and reveal clusters in lower-dimensional space [57]. Overall, machine learning provides powerful tools for predicting material properties, discovering patterns, and optimizing processes pushing the boundaries of functional materials research.

The standard ML training process involves splitting data into training, validation, and test sets, followed by feature extraction and hyperparameter tuning. Figure 1 shows the key steps in training ML models, including data splitting, feature extraction, and hyperparameter tuning, which is critical for optimizing prediction quality.

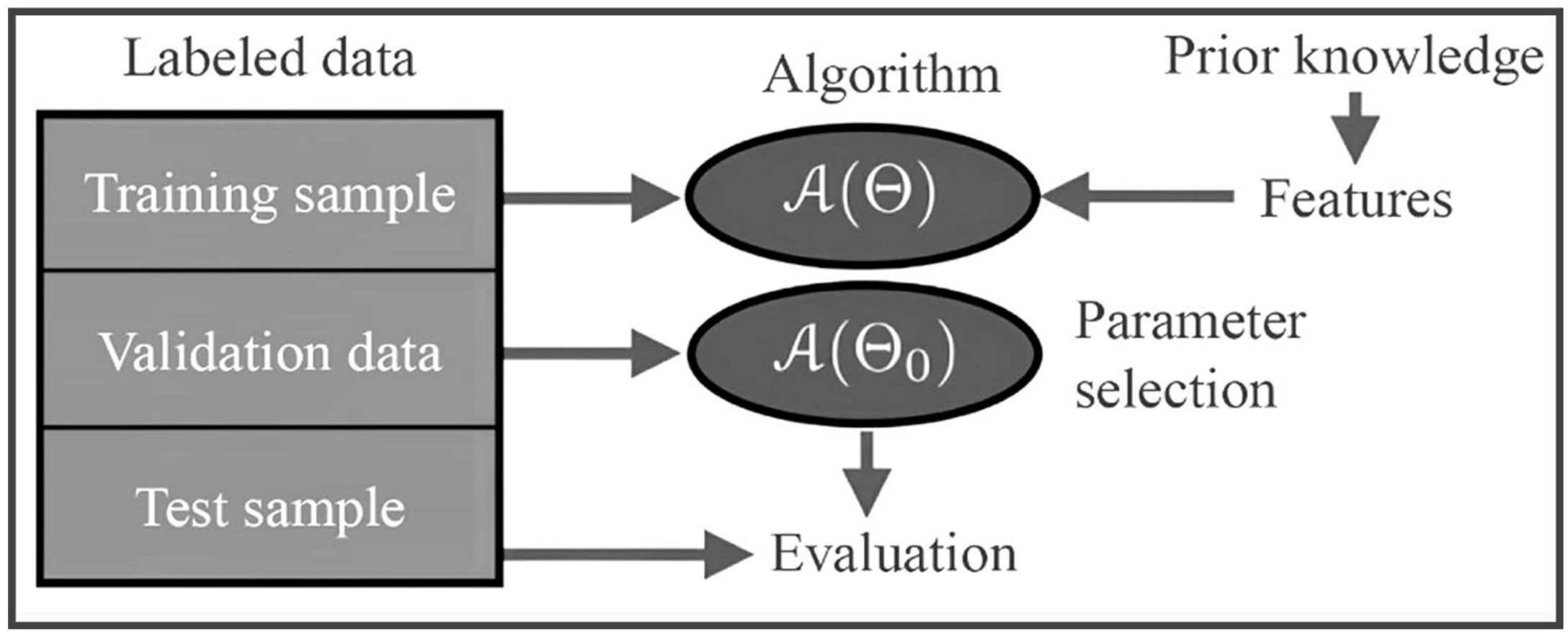

**Figure 1.** A schematic overview of the standard supervised ML workflow used in materials science, illustrating the process of feature selection based on prior domain knowledge, training of algorithm A(Θ) on labeled data, hyperparameter tuning using validation data to obtain optimal model $A(\Theta_0)$, and final model evaluation on independent test samples.

Machine learning in materials science covers a broad spectrum of tasks for analyzing and interpreting different types of data, each with its own preprocessing and analysis methods [58–75]. Structural data capture information about crystalline structures (atomic coordinates, bond types, lattice parameters, symmetry) and are used to predict properties like mechanical, thermal, and electronic behavior. Specialized descriptors, such as coordination number and bond energy are crucial. For example, Ward, L. et al. [58] developed descriptors that accurately predicted the mechanical properties of alloys. Graph Neural Networks (GNNs), which represent atoms as nodes and bonds as edges, have been successfully applied to predict defect formation energy and elastic modulus [5,59]. Additionally, 3D convolutional neural networks (3D-CNNs) classify crystalline structures based on symmetry and lattice parameters [60], while regression methods have been employed to predict thermoelectric properties [61] and stability [62]. Spectral data from techniques like X-ray diffraction (XRD), infrared (IR), and Raman spectroscopy are high-dimensional and noisy, requiring preprocessing methods, such as normalization, smoothing, and noise removal. Oviedo, F.et al. [63] improved the XRD spectral preprocessing to enhance crystalline phase classification. Convolutional neural networks (CNNs) automatically extract key features from spectra. For example, Lee, J.et al. [64] used CNNs for Raman spectrum classification. Autoencoders have also been employed to reduce dimensionality in IR spectra [65], and similar spectral applications include crystalline phase identification via XRD [66] and Raman-based composition prediction [67]. Experimental data such as results from mechanical testing, thermal analysis, and electrochemical measurements often appear as time series or tables and are characterized by noise and time-dependent behaviors. Recurrent neural networks (RNNs) and LSTM networks analyze such data. For instance, McInnes, L. et al. [68] applied LSTM networks to predict material lifetime from degradation data. Regression methods have been used to predict the mechanical properties of composites [69], and ensemble methods like gradient boosting predict thermal properties [62, 70]. Additional applications include predicting battery lifespan using regression approaches [71] and analyzing material corrosion with LSTM networks [72]. Representation learning using autoencoders has proven valuable for extracting key features from experimental data. Figure 2 shows the process of training an autoencoder that helps to reveal hidden features in experimental data, improving the understanding of catalyst structures.

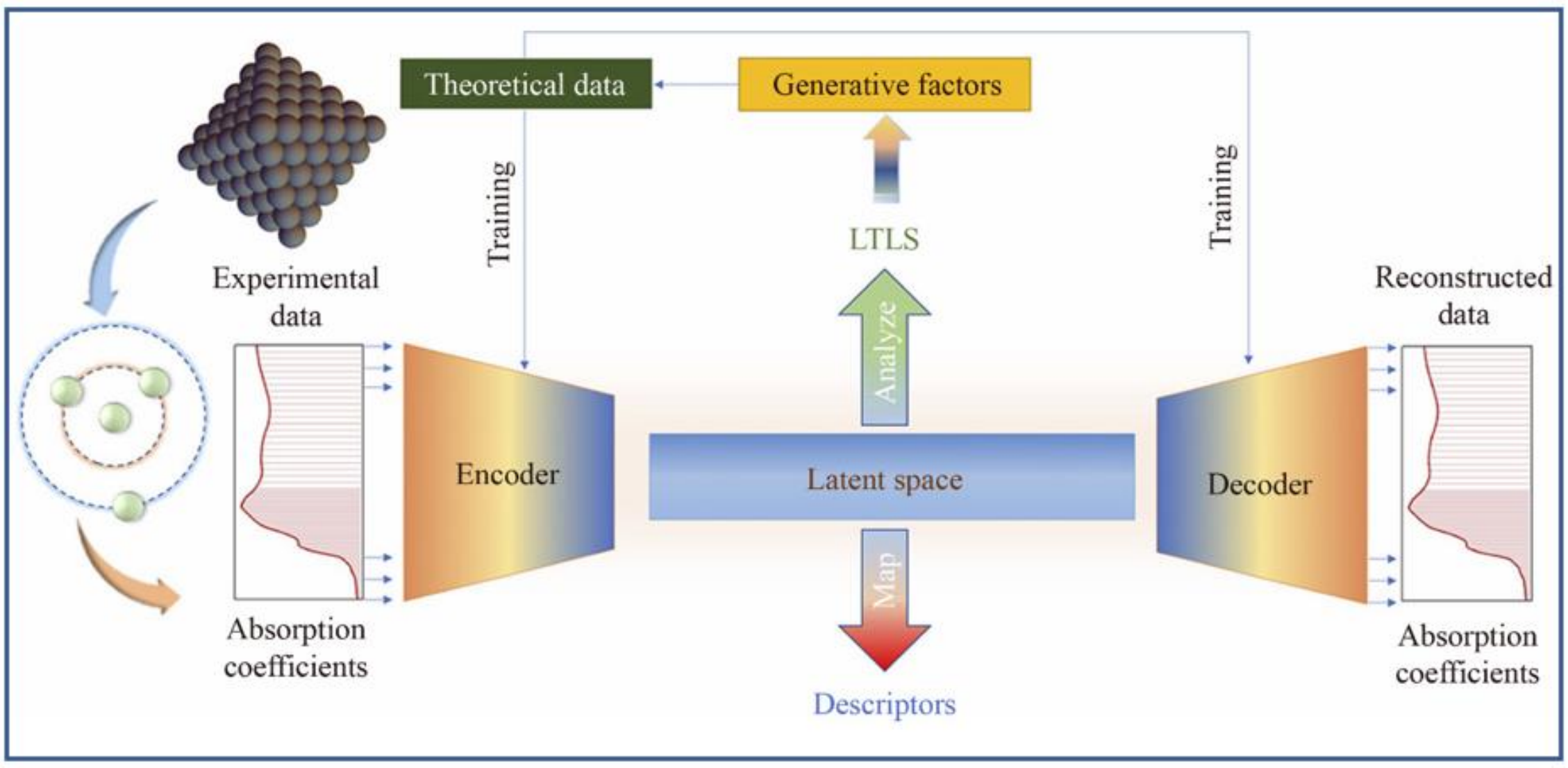

**Figure 2.** Schematic representation of representation learning using an autoencoder for catalyst structural characterization [73].

Modern research often integrates structural, spectral, and experimental data using multimodal learning to enhance prediction accuracy [67,74–77]. AutoML further streamlines model development by automating algorithm selection, hyperparameter tuning, and data preparation [77-79]. In materials science, AutoML is used to predict various material properties, including mechanical, thermal, electrical, and optical characteristics. For example, in these studies [3,78], an AutoML system was successfully implemented to predict the Gibbs energy of various inorganic compounds, significantly reducing the time required for data preparation and model building. The core concept of AutoML involves using algorithms that automatically explore different model configurations, train them, and select the most effective solutions based on predefined quality metrics [79].

One of the key advantages of AutoML is its ability to handle large datasets, which are common in modern materials research. Databases, such as the Materials Project, Open Quantum Materials Database (OQMD), and Automatic Flow for Materials Discovery (AFLOW) contain millions of records on crystalline structures, their properties, and energy characteristics [5,6,80-82]. AutoML enables the use of these large datasets to rapidly develop models capable of accurately predicting the properties of new materials. For instance, in Zhang, L. et al. [8], an AutoML system was employed to determine the electrical conductivity of organic polymers, where the resulting models achieved accuracy comparable to DFT methods but with a significantly shorter computational time. AutoML is also actively used in designing new materials with desired properties. For example, in Zhang, L. et al. [83], AutoML algorithms were utilized to develop new catalysts for efficient water splitting into hydrogen and oxygen via photocatalysis. By leveraging data on the structure and composition of known catalysts, AutoML algorithms were able to propose new materials with enhanced characteristics. Automating model development is particularly crucial when dealing with multicomponent systems, where the number of possible element and structure combinations can reach millions. In studies [9,84], AutoML was applied to optimize the structure of perovskites for solar cells, improving solar energy conversion efficiency by 15% compared to materials selected using traditional methods.

Specialized software platforms such as Auto-sklearn, TPOT (Tree-based Pipeline Optimization Tool), and $H_2O$ AutoML are commonly used for AutoML applications. Studies [11,85,86] provide detailed descriptions of these tools in the context of materials science. For instance, TPOT has been applied for automatic descriptor selection in predicting the hardness of metallic alloys, while Auto-sklearn demonstrated high efficiency in analyzing the thermoelectric properties of semiconductor materials.

One of the promising applications of AutoML is the study of multicomponent alloys, such as high-entropy materials, where the number of possible element combinations is extremely large. In Sutton, A. et al. [87], an AutoML model was developed to predict the phase stability of alloys containing up to five different elements. The automated approach reduced the analysis time by a factor of ten compared to traditional methods. However, despite its advantages, AutoML faces several challenges and limitations. First, model quality heavily depends on the quality and completeness of the input data. Studies [13,14,88,89] indicate that AutoML algorithms do not always handle uncorrelated or insignificant features correctly, which can result in models with low predictive power. Second, many available AutoML platforms require significant computational resources, particularly when dealing with large datasets, which limits their use in environments with restricted computational capabilities [90]. AutoML represents a promising direction in materials science, significantly accelerating the development and optimization of new materials. Automating model selection, hyperparameter tuning, and data preparation makes this tool accessible to a broad range of researchers, enabling the efficient development of highly predictive models with minimal effort. As artificial intelligence technologies continue to advance, AutoML is expected to be increasingly adopted in materials science, facilitating the discovery of new, more efficient, and sustainable materials for energy, electronics, medicine, and other industries [2,91-98].

ML combined with quantum mechanical and molecular mechanic methods (DFT and MD) enables atomic-level modeling of material properties but requires substantial computational resources [96-98]. ML reduces this complexity while maintaining high prediction accuracy. Integrating data representation techniques with generative models offers several benefits: improved noise handling, latent representations that align with design tasks, and automated processes that minimize human error. This unified approach using both discriminative and generative models addresses tasks ranging from property prediction within DFT to inverse design [96-98]. For instance, encoding crystalline structures into latent spaces facilitates the search for materials with desired bandgaps, and generative models such as VAEs and GANs are used to create materials with targeted properties. In Bang, K.et al. [98], VAE generated stable perovskite structures based on DFT data. While analytical methods may suffice for systems with few degrees of freedom, ML becomes essential as dimensionality increases, with discriminative models handling moderate dimensions and generative models reducing complexity to uncover new solutions that meet predefined criteria. For example, GNNs-like Crystal Graph Convolutional Neural Networks predict formation energies by representing crystalline structures as graphs [5,90]. ML also optimizes DFT exchange-correlation functionals [99,100] and supports inverse design through generative models [101,102]. In MD, methods such as DeePMD use neural networks to predict interatomic potentials with DFT accuracy, and active learning facilitates the discovery of new alloys [103-107]. As shown in Table 2, various ML methods are applied for both DFT and MD integration. This combined table illustrates those methods, such as GNN and KRR enhance DFT calculations, while DeePMD and active learning accelerate MD simulations.

**Table 2.** Examples of ML integration with DFT and MD.

| **ML method** | **Application** | **Ref.** |
|---|---|---|
| Graph Neural Networks (GNN) | Prediction of formation energy (DFT) | [96] |
| Kernel Ridge Regression (KRR) | Optimization of exchange-correlation functionals (DFT) | [99,100] |
| Generative Models (GAN, VAE) | Inverse design of materials (DFT and MD) | [101,102] |
| DeePMD | Prediction of interaction potentials (MD) | [103] |
| Active Learning | Discovery of new alloys (MD) | [104] |
| Graph Transformers (Matformer / ComFormer) | DFT-trained crystal property prediction (formation energy, band gap, elasticity) | [94] |
| Diffusion generative models (DiffCSP / MatterGen) | Inverse design & crystal generation with DFT relaxation | [24] |
| Autonomous lab + active learning | Closed-loop ML-guided synthesis integrated with computation and robotics | [48] |

This table shows how different ML methods are utilized to improve both DFT and MD processes. GNNs and KRR methods are used to enhance the accuracy of DFT calculations by predicting formation energies and optimizing functionals. Meanwhile, methods like DeePMD and active learning are applied in MD to accelerate simulations and aid in discovering new alloys. Despite these advances, challenges remain regarding data quality, model interpretability, and the integration of models across different scales. Future research will focus on enhancing data quality, developing explainable AI methods, and creating hybrid models that merge ML with fundamental physical principles.

## 3. ML methods for property prediction

Modern materials science uses ML to predict mechanical, thermal, electrical, and optical properties. Traditional approaches, such as computational quantum chemistry and MD simulations require substantial time and resources, particularly for complex systems. ML reduces this burden by uncovering hidden patterns in large datasets, making predictions faster and more efficient. Material properties (hardness, melting temperature, ionic conductivity) are studied at both macroscopic and microscopic levels through computational modeling and experimental measurements. Building accurate models to link material structure and properties is challenging due to unknown dependencies, and unsatisfactory experimental results can waste significant resources [83]. Therefore, intelligent predictive models are needed to minimize time and computational costs [6,18,92]. ML analyzes large datasets to identify complex nonlinear relationships between structure and properties [7]. Advanced ML models such as deep neural networks and gradient boosting have notably improved prediction accuracy [15,27,46,90]. As shown in Figure 3, ML in materials science involves three main stages: data processing (including feature extraction), model training to establish structure-property relationships, and using the trained model to predict properties of new materials.

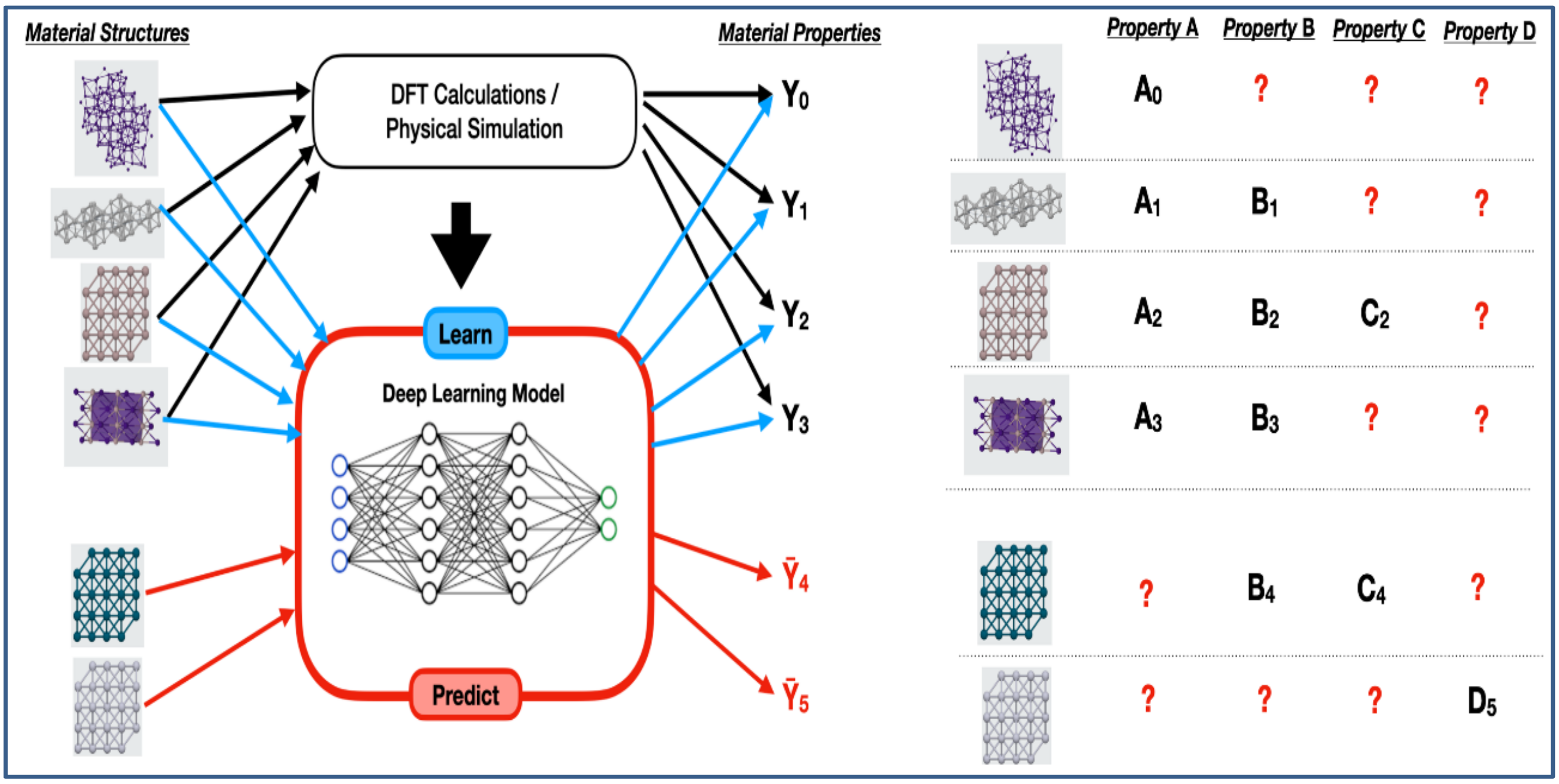

**Figure 3.** A typical workflow of ML-based property prediction: from structural/chemical data preprocessing and feature engineering, to training of regression models, and their application in predicting target properties such as bandgap or hardness.

Figure 3 shows the overall ML process, highlighting how data processing, model training, and prediction reduce time and resources compared to traditional methods. This shows how material structures (for example, obtained from DFT calculations or other physical simulations) can be used to train a deep learning model that predicts various material properties. Once the model is trained, it can take a new material's structure as input and output predictions for properties of interest (Property A, B, C, D), thus speeding up the process of material discovery and development. For example, the Property-Labeled Materials Fragment (PLMF) method by Isayev et al. [50] uses gradient boosting to classify materials (metals vs. insulators) and predict properties like bandgap, bulk modulus, shear modulus, Debye temperature, heat capacities, and thermal expansion. Prediction accuracy is evaluated with metrics such as ROC curves, RMSE, MAE, and $R^2$ [41,47,93]. ML methods cover a wide range of applications. GNNs predict electronic and mechanical properties [45,62,99]. They model atomic structures as graphs, where atoms are represented as nodes and interatomic bonds as edges. Through multiple message-passing layers, each node aggregates information from its neighbors, allowing the model to capture both local and global structural features. This hierarchical representation enables the prediction of properties such as formation energy, elastic constants, and bandgap by learning structure–property relationships directly from graph topology [5,15,21,43,45,62,90]. GANs aid in designing new crystalline structures [101,102], and active learning optimizes the discovery process by reducing experiments [81]. Beyond supervised pipelines for property prediction, generative models and active learning act as complementary levers for discovery. GAN/VAEs provide prior-guided sampling in composition/structure spaces to propose chemically plausible candidates, while active learning closes the loop by selectively querying the next most informative calculations or experiments (uncertainty-, diversity-, or goal-oriented acquisition). Combined, a generative prior proposes candidates and the acquisition policy prioritizes which to evaluate next, thereby reducing data-collection cost and accelerating convergence to target properties. The GNN framework models a material's crystal structure as a graph, where nodes correspond to atoms and edges represent bonds or distances. These models use message passing mechanisms to propagate information across the structure, enabling accurate predictions of formation energy, bandgap, elasticity, and more. Figure 4 illustrates a typical GNN architecture employed in material property prediction.

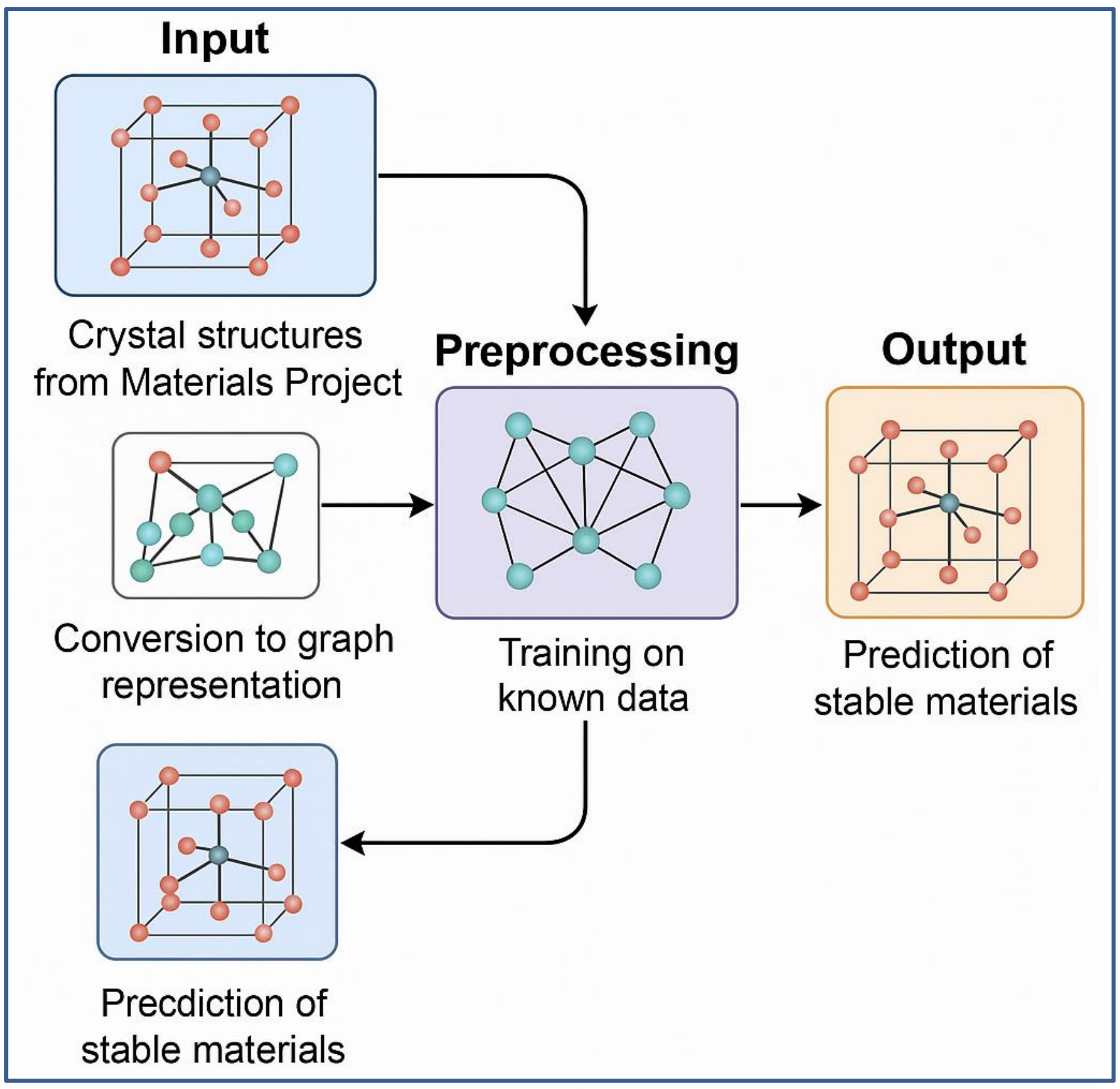

**Figure 4**. Workflow of GNN-based prediction of stable materials.

Thus, ML accelerates material development by enabling both macroscopic (mechanical, electrical, thermal) and microscopic (binding energy, lattice parameters) property predictions [12,84].

For mechanical properties, ML models, such as neural networks and gradient boosting can predict characteristics (hardness, tensile strength, elastic modulus) based on composition and crystalline structure [46]. These models rely on a wide range of features describing elemental, structural, thermodynamic, and spectral characteristics of materials (Table 3). GNNs, for instance, model the relationship between atomic interactions and macroscopic properties, predicting Young's modulus with accuracy comparable to DFT [47].

**Table 3**. Common features used in ML-based property prediction of materials

| FEATURE CATEGORY | EXAMPLES OF FEATURES | TYPICAL USE CASES | REF. |
|---|---|---|---|
| **Elemental Descriptors** | Atomic number, electronegativity, atomic radius, ionization energy, valence electrons | Prediction from composition using RF, GBT | [41, 45, 49] |
| **Structural Features** | Lattice constants, symmetry, packing density, coordination number | Hardness, bandgap, phase stability | [5, 50, 58] |
| **Electronic Properties** | Density of states, Fermi level, band edge positions | Conductivity, bandgap | [84, 93, 99] |
| **Thermodynamic Features** | Formation energy, melting point, specific heat, Gibbs energy | Stability, decomposition prediction | [6, 18, 92] |
| **Spectral Descriptors** | Peak intensities and positions in IR, Raman, XRD spectra | Material classification and phase identification | [64-73] |
| **Process-Related Features** | Synthesis temperature, cooling rate, annealing time | Steel hardenability, catalyst optimization | [108-110] |
| **Statistical/ Derived** | Feature mean, std, SHAP value, feature correlation | Feature importance, interpretability (XAI) | [47,102-104] |
| Graph-based descriptors | Node/edge/topological features from crystal graphs (coordination, bond lengths, Voronoi indices) | GNN-based property prediction; interatomic potentials | [97] |
| Text-mined descriptors | NLP-extracted terms from literature (e.g., activation energies, site occupancy, synthesis conditions) | Knowledge-guided features; low-data regimes | [108] |
| Latent descriptors (VAE/GAN/diffusion) | Low-dimensional embeddings capturing structure–property relations | Inverse design; candidate ranking | [21] |

Additionally, combining ML with MD has enabled the simulation of plastic deformation processes in nanomaterials [108]. For thermal properties, ensemble methods like random forests and deep CNNs have proven efficient in predicting thermal conductivity and expansion, while hybrid models integrate ML with physicochemical models for better interpretability [109,21].

ML is also used for predicting electrical properties such as conductivity, dielectric permittivity, and bandgap. Deep learning methods, including GCNs, offer higher accuracy in predicting optical and electronic properties than traditional models [111]. Autoencoders model material absorption spectra, aiding organic semiconductor development [112–117]. For example, generative models have been applied to predict the optical properties of quantum dots, as shown in Figure 5**.**

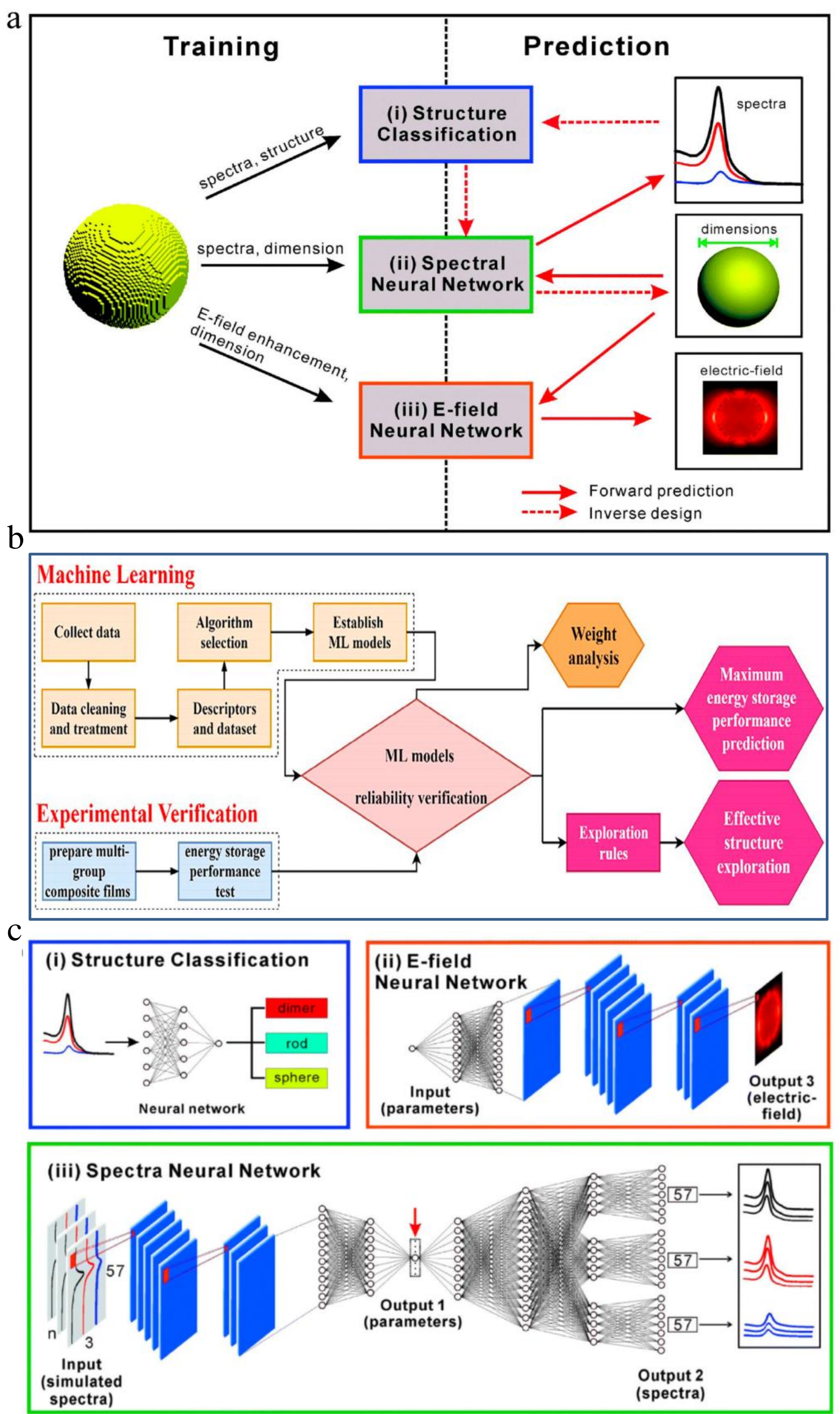

**Figure 5.** ML pipelines and model architectures for optical property tasks: (a) dataset curation and preprocessing; (b) end to end workflow for property prediction and inverse design (data → model selection/training → validation → experimental verification); (c) detailed architectures of the models in (b) including inputs hidden blocks and outputs for (i) structure classification (ii) spectrum prediction and (iii) E field prediction.

Figure 5 presents ML architectures for forward prediction and inverse design in photonics. Figure 5a outlines dataset curation and preprocessing. Figure 5b illustrates the end to end workflow from data and model selection and training to validation and experimental verification. Figure 5c details three model architectures used in b including (i) structure classification, (ii) spectrum prediction, and (iii) E field prediction, with inputs, hidden blocks, and outputs indicated. This consistent pipeline ensures robust generalization and efficient exploration of the design space.

Figure 6 summarizes a typical feature engineering workflow for materials science including data integration and cleaning feature extraction and selection and model training [114]. ML is applied in predicting metallic alloy properties using CNNs on microstructure images [117], designing high-entropy alloys through active learning [118], optimizing catalysts with GNNs [119], and predicting perovskite solar cell stability [120]. AutoML has been effectively used for polymer property prediction by automating hyperparameter optimization [121-128]. However, training complex ML models on large datasets requires significant computational resources.

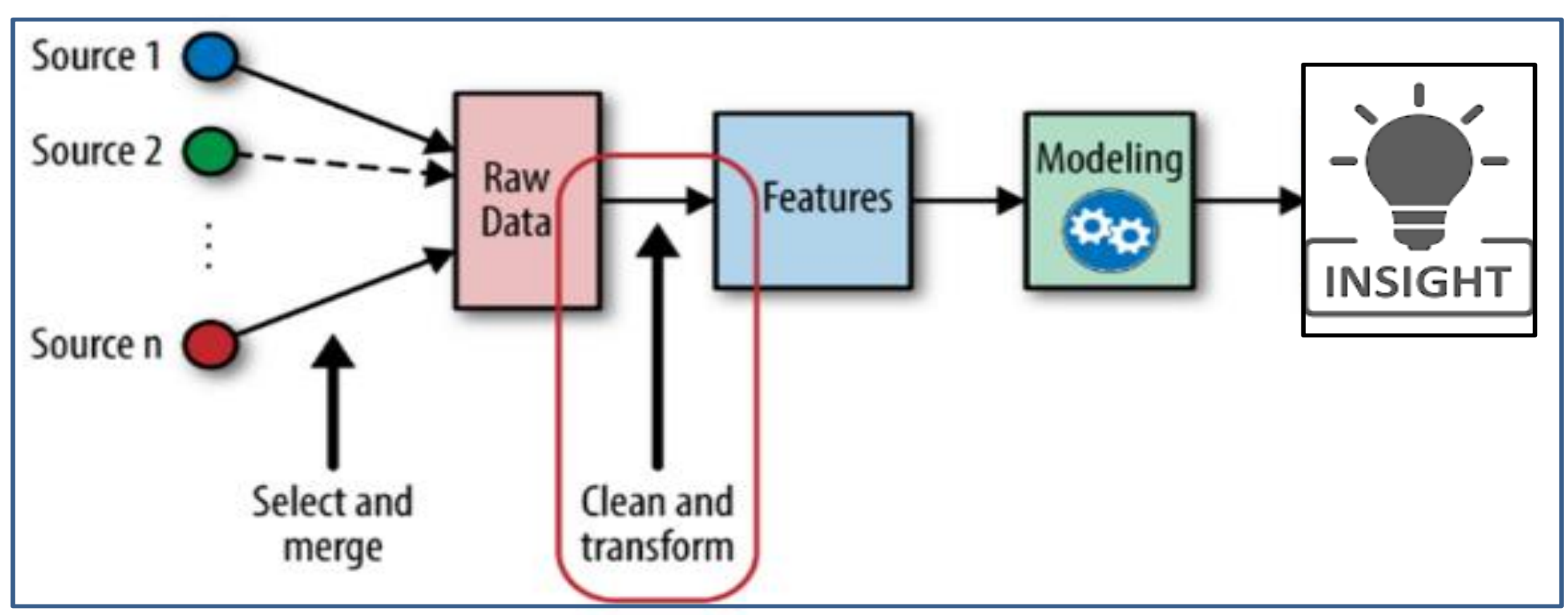

**Figure 6.** Feature engineering pipeline for materials informatics showing data integration and cleaning feature extraction and selection and model training and validation.

Heterogeneous structure–property relationships often violate single-model assumptions. Divide-and-conquer modeling addresses this by splitting regimes or samples and training tailored learners per partition: stage-wise models improved glass-transition onset predictions in chalcogenides [122, 123], and sample-level clustering with cluster-specific models boosted creep-life prediction for Ni-based single-crystal superalloys [124]. Extending this idea, descriptors divide-and-conquer integrates rough-set feature mining with rule extraction to achieve both high accuracy and interpretable structure–activity rules across diverse datasets [125]. In practice correlation can be assessed with Pearson or Spearman metrics and feature importance can be estimated with SHAP or permutation importance to identify high impact descriptors and reduce redundancy however these analyses are part of standard workflows and are not depicted in Figure 6.

Machine learning in prediction and material retrieval often faces limitations such as lack of data, uneven data distribution, noise, and missing values. Materials ML faces three tensions: high dimension versus small sample, accuracy versus usability, and learned patterns versus domain knowledge. These tensions motivate pipelines that embed domain knowledge [129, 130]. Expert-weighted multi-layer feature selection and correlation-aware pruning improve stability and generalization, and Auto-MatRegressor combines meta-learning with AutoML to recommend near-optimal models for a given dataset [9,13,18]. Databases like Materials Project, OQMD, AFLOW, and NOMAD primarily contain inorganic crystalline data, leaving amorphous materials, polymers, composites, and defective materials underrepresented [113]. Literature-aware pipelines complement curated databases. NLP-based descriptor recognizers mine latent descriptors from texts, such as activation energies in solid electrolytes [108]. Abstract-level knowledge discovery frameworks enrich interpretable structure–property insights, for example in Ni-based single-crystal superalloys [113]. A semantic-features-fused NER improves entity extraction with reduced corpus requirements and aids cross-domain generalization [116]. Standardized construction of high-quality corpora for materials text mining further improves downstream model accuracy [32].

Synthetic data generation using GANs and active learning is being developed to enhance these datasets [123]. Modern deep neural networks often act as “black boxes,” complicating the understanding of underlying physical mechanisms. Explainable AI (XAI) methods (SHAP, LIME, Integrated Gradients) are under development to clarify feature contributions [124]. Hybrid models that combine physical laws with ML such as integrating DFT equations with neural networks offer more interpretable predictions [125]. Dimensionality reduction (PCA, autoencoders), distributed computing, quantum algorithms [126], and transfer learning [127] are also used to reduce computational costs. Another challenge is bridging different scales from atomic to macroscopic properties. ML helps develop multiscale models that integrate data from various sources, such as predicting interatomic potentials to accelerate MD simulations [128]. However, integrating multi-scale data remains complex. Experimental validation of ML predictions is also resource-intensive, prompting the development of autonomous laboratories and robotic platforms to synthesize and test materials automatically [129]. Additional challenges include intellectual property, data accessibility, and AI ethics, which require interdisciplinary solutions and standards [130]. Quantum computing, with Quantum ML (QML), offers new possibilities for modeling complex materials (high-temperature superconductors, quantum dots) by accelerating calculations and improving accuracy [131]. Advances in experimental data processing like X-ray diffraction and electron microscopy are leading to more precise ML models [132]. ML is crucial for addressing global challenges, such as climate change and the energy transition, by developing new catalysts for $CO_2$ capture, materials for hydrogen energy, and high-energy-density batteries [133]. Additive manufacturing combined with ML enables the creation of materials with tailored properties for various industries [134]. The future of materials science and molecular research is increasingly linked to ML, which accelerates the discovery and optimization of new materials. Although modern AI methods are widely used for material property prediction, their full potential has yet to be realized [134,135]. Deeper integration of ML in research and development is expected to drive revolutionary changes in functional materials, nanotechnology, and bioengineering.

## 4. Design of new materials

In modern materials science, developing steels with tailored properties, such as hardenability and hardness is a major challenge. Traditional methods for selecting steel compositions are time-consuming and costly. Advances in computational technologies, especially ML, now enable faster and more accurate predictions of material properties based on chemical composition and processing conditions [136-139]. One critical task is designing steel compositions to achieve the desired hardenability, which directly affects mechanical properties like wear resistance and durability. For instance, Tomacich et al. [136] proposed an innovative approach using artificial neural networks (ANNs) to predict chemical compositions based on the desired Jominy hardenability curve. Their model, trained on databases of hardness, microstructure, and composition, optimizes the balance between cost and required hardenability. Trzaska and Sitek [137] introduced a hybrid method combining ANNs and genetic algorithms (GAs) to calculate steel composition for achieving specific hardness after cooling from the austenitization temperature. They built a database of 550 continuous cooling transformation (CCT) diagrams and developed a hardness model using ANNs, with GAs determining the optimal composition. The high accuracy of this method is confirmed by comparing predicted and experimental hardness values. Li et al. [138] developed a combined ML model using random forest for classification and k-nearest neighbors/random forest for regression to predict hardenability curves for boron-free steels, outperforming conventional methods. Gemechu et al. [139] applied regression neural networks optimized via Bayesian methods to predict steel hardness, achieving high accuracy as demonstrated by low RMSE and high $R^2$ values (Table 4).

**Table 4.** Statistical values for evaluating the hardness model [139].

| Dataset | Mean absolute error, HV | Standard deviation of the error, HV | Ratio of standard deviations | Pearson correlation coefficient |
|---|---|---|---|---|
| Training | 30.9 | 44.3 | 0.27 | 0.96 |
| Validating | 33.6 | 46.4 | 0.28 | 0.96 |
| Testing | 33.7 | 50.1 | 0.30 | 0.95 |
| Verifying | 32.7 | 39.0 | 0.29 | 0.95 |

A study demonstrates the potential of machine learning (ML) for predicting steel properties and optimizing heat treatment processes. Zhang, Y. et al. [140] presented a strategy for designing high-manganese TWIP steels using ML, incorporating comparative modeling, SHAP analysis, and multi-objective optimization to develop alloys with enhanced strength and ductility. In the study by Schmidt et al. [141], ML methods based on graph neural networks were applied to predict the behavior of steels under real operating conditions. Patel, A. et al [142] utilized convolutional neural networks for automated steel design with targeted mechanical properties. Wang, C. et al. [143] demonstrated the use of Generative Adversarial Networks to develop new alloys with high strength and corrosion resistance. Generative ML models (VAE, GANs) are increasingly used in material design, predicting the physical and functional properties of new compounds [144]. In solar energy applications, GANs and VAEs are employed to identify stable perovskites with enhanced resistance to moisture and thermal stress [145]. Wang, Y. et al. [146] applied GANs to generate stable perovskite structures with optimized optical properties. In photocatalysis, GANs accelerate the discovery of new materials by identifying stable semiconductor compounds with a narrow bandgap suitable for visible-light-driven catalysis [147]. Metal-organic frameworks (MOFs), known for their high porosity and tunable chemical structures, are widely applied in gas storage and water purification. GANs and VAEs significantly expedite MOF discovery by predicting promising structures with high adsorption capacity, as demonstrated in the study by Patel et al. [148]. Quantum dots, widely used in displays and biomedical sensors, are designed using VAEs to predict materials with maximum quantum efficiency and photostability [149]. GANs are also applied in developing new organic luminescent compounds, such as OLED materials with optimized emission spectra [150-153]. For molecular design, latent-space models and adversarial training have yielded property-aware priors that help propose synthetically accessible candidates, complementing graph-based predictors and reinforcement learning used for goal-directed optimization [4]. In electrochemistry, GANs aid in the discovery of high-capacity and stable cathode and anode materials for batteries [153]. Despite numerous successful applications, challenges remain in the use of generative models in materials science, including the lack of high-quality experimental data and model interpretability. Addressing these issues requires integrating ML with physicochemical simulations and quantum mechanical modeling [152,153].

In summary, conventional regression-based ML methods are still dominant in steel design, while emerging generative models show promise in discovering novel alloy compositions. However, challenges remain in model interpretability and experimental validation. To summarize the capabilities and limitations of the major ML and DL approaches used for property prediction, we present a comparative overview in **Table 5**. This includes their typical input features, level of interpretability, prediction accuracy, and selected application domains.

**Table 5.** Comparison of ML, DL and XAI approaches in materials property prediction tasks (trade-off between interpretability and accuracy)

| **Method** | **Features used** | **Explainability** | **Accuracy** | **Application Ex.** | **Ref.** |
|---|---|---|---|---|---|
| Graph Transformers (Matformer/ComFormer) | Crystal graphs with periodic invariance; geometric distances/angles | Medium | Very High | Formation energy, band gap, elasticity (Matbench) | [37, 94] |
| Language-model descriptors (text-based Transformers) | Text descriptions of composition/structure; expert rationales | High | Medium–High | Property prediction with human-aligned explanations | [38] |
| RF | Composition, elemental descriptors | Medium | High | Bandgap, hardness prediction | [49,84] |
| E(3)-equivariant GNNs (NequIP-class) | Local environments with rotational/translation equivariance | Medium | High | Accurate interatomic potentials; MD surrogates | [97] |
| Gradient Boosting Trees (GBT) | Composition, structure-based features | Medium | High | Elastic modulus, formation energy | [92,153] |
| DNN | Raw input, feature-engineered vectors | Low | Very High | Thermoelectric materials | [115] |
| GNN | Graph structure of crystals | Low | Very High | Formation energy, bandgap | [5,45, 96,99] |
| SHAP+ RF | Composition, structure | High | High | Feature attribution for bandgap | [142] |
| AutoML + XAI | Auto-selected + SHAP | High | High | Polymer conductivity | [107] |

As shown in Table 5, the selection of an appropriate ML or DL method in materials science depends not only on prediction accuracy but also on the trade-off between interpretability and complexity. Traditional ensemble models such as Random Forest and Gradient Boosting are still widely used due to their robustness and relatively good transparency. On the other hand, deep learning architectures like GNNs and DNNs demonstrate superior performance on large and complex datasets, albeit at the cost of reduced explainability. Recent developments in explainable AI (XAI), including SHAP and LIME, provide new avenues for enhancing model transparency without significantly sacrificing accuracy. This is especially crucial in applications where understanding feature contributions is essential, such as in alloy design or failure prediction in structural materials. Moreover, AutoML frameworks incorporating XAI are emerging as powerful tools for accelerating materials discovery by automatically selecting optimal features and model architectures, offering a promising direction for high-throughput screening and data-driven materials design.

Overall, the synergy between advanced ML models, domain knowledge, and explainability tools is paving the way toward more intelligent, interpretable, and efficient materials development pipelines.

## 5. ML and discovery of new materials

The search for new materials with high-performance characteristics is a key challenge in materials science. Currently, experimental and computational screening methods for new materials rely on element substitution and structural transformation; however, the search space for composition, structure, or both parameters is typically highly constrained [154]. Additionally, these methods require significant computational or experimental resources and often lead efforts in the wrong direction within an exhaustive search framework, resulting in substantial time and resource expenditures. Given these limitations and the advantages of machine learning, a fully adaptive method has been proposed that integrates ML with computational modeling to evaluate and select new materials in silico, allowing for more efficient discovery of promising materials. The general process of machine learning for material discovery is illustrated in Figure 7, where the ML system consists of two main components: the training system and the prediction system.

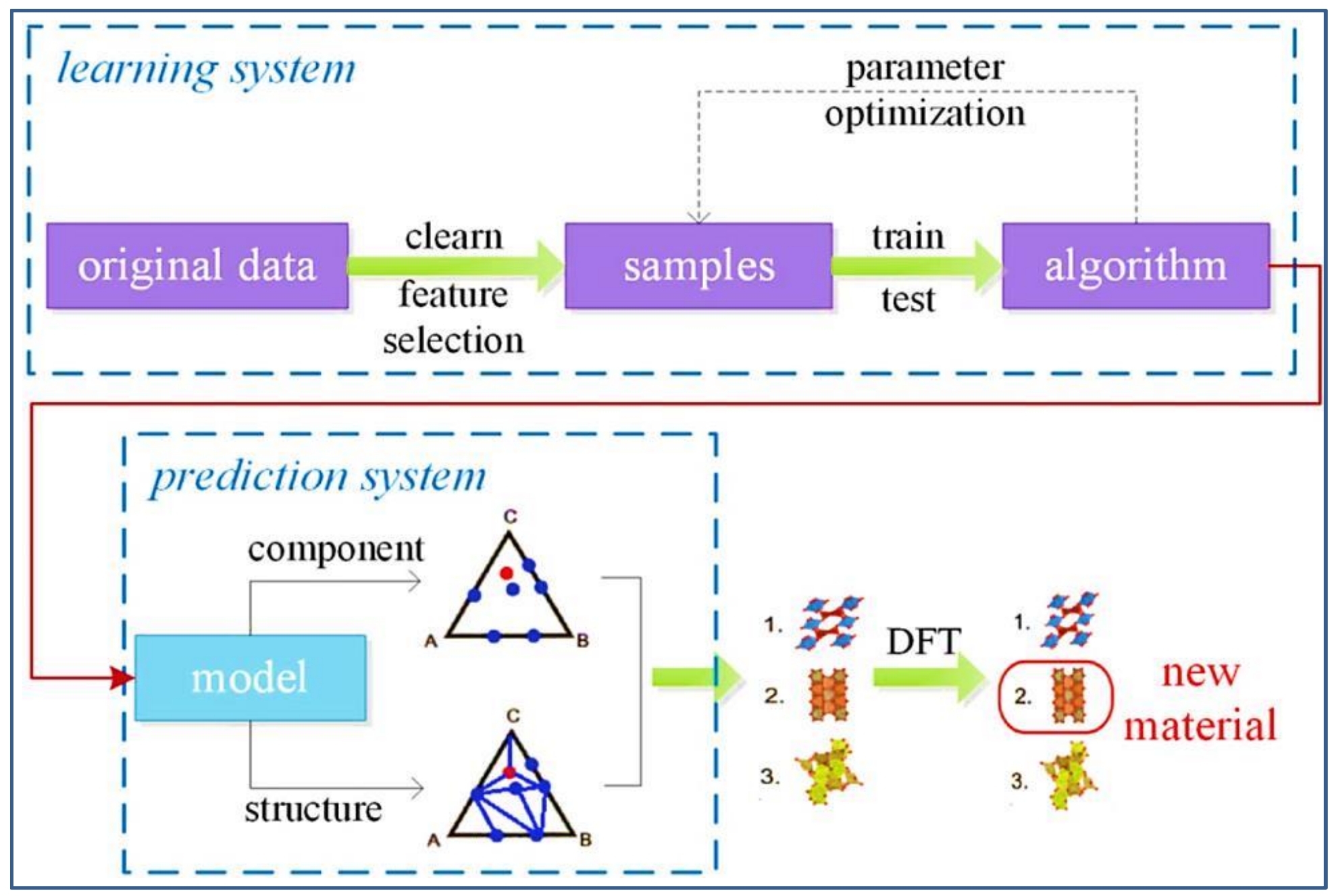

**Figure 7.** General workflow of ML-based material discovery: the upper pipeline represents data processing, feature extraction, and model training, while the lower pipeline applies the trained model for prediction and validation of novel materials via first-principles simulations. This two-stage approach streamlines discovery by integrating ML predictions with DFT calculations [155].

Figure 7 illustrates a two-stage workflow for materials discovery. The upper "learning system" cleans and processes the original data, applies feature selection, and uses the resulting samples to train and optimize a machine learning algorithm. The lower "prediction system" then leverages the trained model to propose new material compositions or structures, which are subsequently validated using DFT calculations. By integrating ML-based predictions with first-principles simulations, this approach accelerates the search for new materials with desired properties, significantly reducing the time and cost associated with trial-and-error experimentation.

New materials are typically predicted using a recommendation and testing approach, where the prediction system selects candidate structures based on recommended compositions and structures, followed by Density Functional Theory (DFT) calculations to evaluate their relative stability. Currently, various machine learning methods are employed to search for new materials with high-performance characteristics (Table 6). These methods can be broadly classified into approaches aimed at predicting crystalline structures and those focused on composition prediction, which will be discussed in detail in subsequent sections. However, in Farrusseng, D. et al. [156], an attempt to combine Artificial Neural Networks (ANN) and Genetic Algorithms (GA) did not accelerate the discovery of new materials due to insufficient awareness of key descriptors. These approaches demonstrate that ML can significantly accelerate material discovery by reducing the need for exhaustive experimental and computational efforts.

**Table 6.** Applications of machine learning in the discovery of new materials.

| Application description | ML method | Achievement | Ref. |
|---|---|---|---|
| Generative inverse design | Diffusion-based crystal generator (MatterGen) | Stable, diverse inorganic structures conditioned on target property | [4] |
| Autonomous closed-loop synthesis | A-Lab robotics + active learning | 41/58 targets realized in 17 days across oxide/phosphate families | [48] |
| The design of new guanidinium ionic liquids | ANN | 6 new guanidinium salts | [157] |
| Finding nature's missing ternary oxide compounds | Bayesian | 209 new compounds | [158] |
| Discovery of new compounds with ionic substitutions | Bayesian | Substitution Rates of 20 Common Ions | [159] |
| Discovering crystals | DBSCAN & OPTICS | Acceleration of finding new materials | [160] |
| Screening new materials in an unconstrained composition space | Bayesian | 4500 new stable materials | [154] |
| ML-assisted materials discovery using failed experiments | SVM | A success rate of 89% | [161] |
| Virtual screening of materials | ANN | Failed | [156] |
| Large-scale AI-driven crystal discovery | GNN scaling (GNoME) + high-throughput DFT | 2.2M candidates; ~381k stable structures on convex hull | [162] |

ML has revolutionized the discovery of new materials by identifying patterns in large datasets and significantly reducing the need for costly experiments. Traditional quantum chemistry and Density Functional Theory (DFT) methods are limited by high computational demands and the complexity of configurational space [163]. Before the 1980s, predicting new material structures was considered infeasible and described as "one of the shortcomings of physics" [164]. In recent decades, ML has been actively applied to this problem. Curtarolo et al. [165] combined heuristic algorithms with quantum mechanical calculations to discover new binary alloys, confirming their thermodynamic stability via DFT, though their approach was restricted to known structures. Ceder et al. [166] showed that relationships between electronegativity, atomic size, and spatial arrangement could aid in material prediction. Fischer et al. [167] developed the Data Mining Structure Predictor (DMSP) method, directing quantum mechanical calculations toward the most promising materials. Phillips and Voth [168] introduced DBSCAN and OPTICS clustering algorithms to identify new compounds in large datasets.

Further ML advancements have led to systematic material prediction. Liu et al. [169] introduced a model combining random data generation, feature selection, and classification algorithms, predicting new Fe-Ga compounds while reducing computation time by 80%. In 2016, Roekeghem et al. [170] applied ML to analyze 400 new perovskite materials, identifying 36 stable compounds at high temperatures. ML has also transformed the discovery of organic light-emitting diodes (OLEDs). Rafael et al. [16, 171] used neural networks to narrow down 400000 candidate materials to 2,500, leading to OLEDs with over 22% efficiency. Similarly, Sendek et al. [172] applied logistic regression to lithium-ion conductors, reducing 12831 compounds to 317, ultimately identifying 21 promising materials. Analyzing the failed experiments also provides valuable insights into material synthesis boundaries. Studies [161,173] showed that ML could predict inorganic compound formation conditions with 89% accuracy.

Predicting chemical composition is another crucial aspect of material discovery. Hautier et al. [158] used Bayesian statistical methods to analyze 183 oxides, predicting 209 new ternary compounds while reducing computational costs by a factor of 30. Meredig et al. [154] expanded this approach, predicting 4,500 new stable compounds and reducing computational time by six orders of magnitude. Another key application is ionic substitution prediction. Hautier et al. [159] developed a method to evaluate the probability of ion substitutions, leading to the discovery of promising quaternary compounds. As shown in Figure 8b, positive values indicate likely substitutions, while negative values suggest unlikely replacements, opening pathways for synthesizing unconventional materials. Figure 8a demonstrates how the chemical similarity of different elements, quantified by their Mendeleev number, can be correlated to the number of new compounds formed by combining those elements in an *A*-*B*-O system [158]. This provides insights into utilizing elemental properties and ML predictions to guide the exploration of new material compositions. Similarly, Figure 8b shows how the gab correlation metric captures ionic radii differences between common ICSD cations [159]. While not specific to ceramic electrochemical cell (CEC) electrolytes or electrodes, such descriptors of elemental properties and structural features could also inform ML models for predicting promising compositions and structures when designing novel materials for CECs. The concepts are generally applicable, but care must be taken to use descriptors and training data relevant to the target CEC materials to produce valid predictions. Validated on 2,967 quaternary ionic compounds from the ICSD database, the model's substitution rules facilitate exploration in this relatively unexplored field. Creep resistance is crucial for nickel superalloys, and Shi et al. [174] developed the DCSA method, which classifies creep mechanisms using K-Means clustering and optimizes ML models (RF, SVR, GPR, LR, RR), successfully predicting service life.

High-entropy alloys (HEAs), composed of ≥5 elements (5-35 at.%), exhibit excellent mechanical properties and typically form FCC and BCC phases [175-178]. Zhang et al. [179] used genetic algorithms (GA) to optimize ML models, achieving 88.7% classification accuracy and 91.3% phase prediction accuracy. Dai et al. [180] improved predictions using low-dimensional feature selection, though larger datasets are needed. Wang et al. [181] developed MIPHA for microstructure analysis, while Huang et al. [182] used ML models (BP ANN, RF, Bagging) for accurate TTT diagram prediction in stainless steel. Wang et al. [183] applied RF to predict RAFM steel yield strength and elongation with high accuracy.

Lithium-ion batteries involve complex performance variables [184]. Multi scale modeling from atomistic to device and pack levels has long underpinned rational battery design and is increasingly combined with ML and high throughput computing [177, 178]. Li-S batteries offer a high theoretical specific energy (2,567 Wh/kg) but suffer from rapid degradation over time due to issues such as the polysulfide shuttle effect [185-187]. To address these challenges, earlier surveys have outlined standard machine learning (ML) procedures for materials design and highlighted key gaps and opportunities specifically for such complex electrochemical systems [155, 187]. Coupling ab initio data with ML has further elucidated the chemical factors controlling reaction kinetics in battery systems, such as the influence of material disorder and solvent effects [188]. Kilic et al. [189] used ML to analyze discharge capacity and cycle life, highlighting the role of carbon structures and encapsulated cathodes. Shandiz et al. [190] found that cathode crystalline structure significantly impacts battery properties, with RF and extremely randomized trees achieving the highest accuracy.

Perovskite solar cells, while cost-effective, face stability and toxicity challenges [191-193]. Wu et al. [194] applied ML and DFT to screen 230,808 HOIP candidates, identifying 132 promising materials. Another study on 404 perovskite elements revealed that mixed cations, multi-spin coating, and low-humidity storage enhance stability [195], though the lack of standardized testing limits ML applications. Metallic glass, known for its unique mechanical properties [196-199], is difficult to analyze using traditional methods.

Banadaki et al. [200] used the PPM method with HDBSCAN clustering to classify short-range order into 30 groups. Xiong et al. [201, 202] applied RF to predict glass-forming ability and elastic moduli based on 6,471 alloys, using features like total electronegativity (TEN), average atomic volume (VA), and atomic size difference ($\delta$). Statistical radius (rs) proved the most effective for atomic size predictions, and SVR with the Pearson VII universal kernel function (PUK) delivered the highest accuracy. The CVE values of the SVR-PUK-TGS and SVR-PUK-TKS models were small (6.0417 for K and 2.0648 for G), indicating high prediction accuracy of these properties in BMG. Analysis revealed that glass-forming ability is enhanced by high mixing entropy, thermal conductivity, and negative mixing enthalpy ($\sim -28$ kJ/mol) [201].

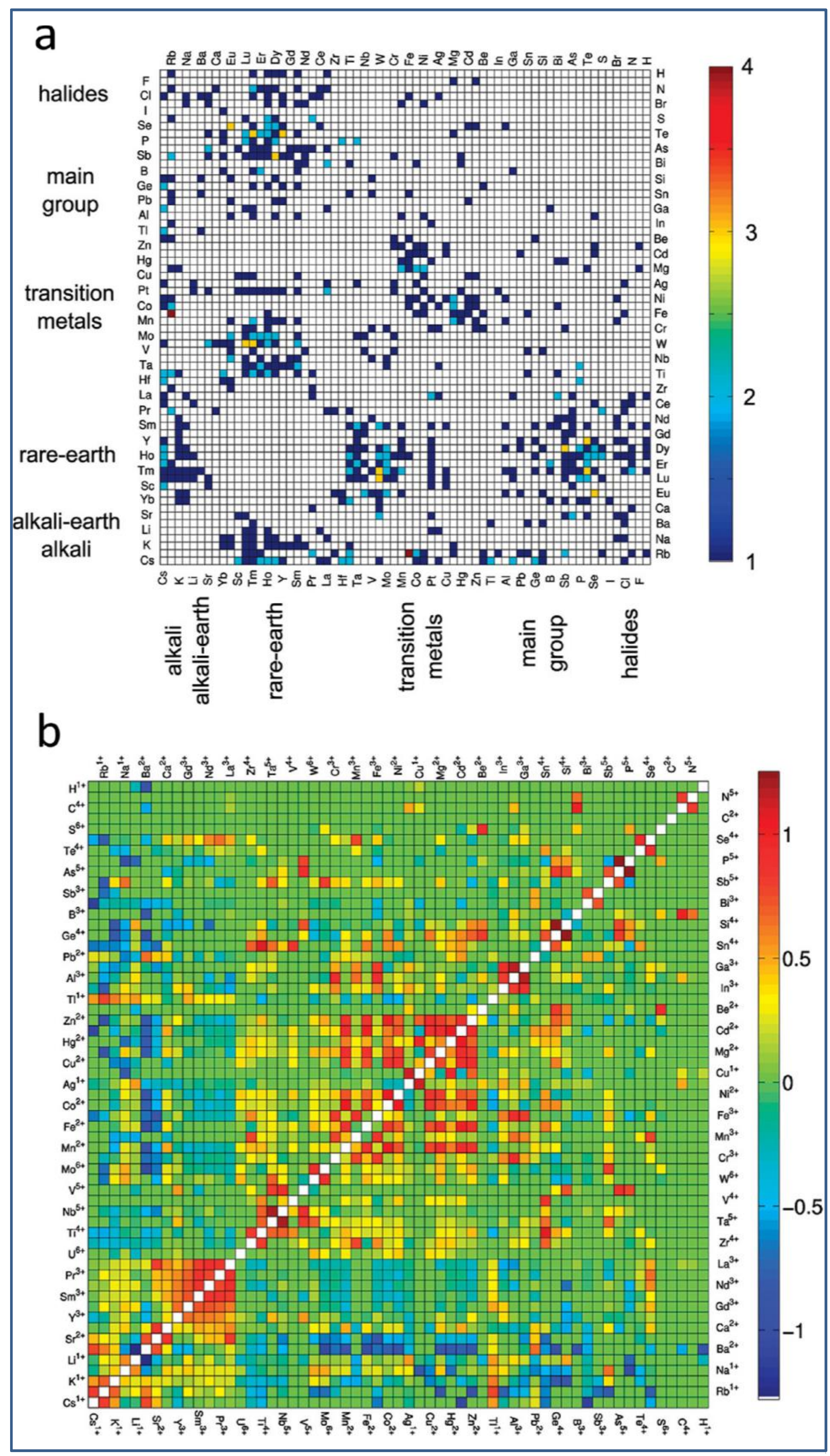

**Figure 8.** Landscape of the A–B–O compound space: (a) Distribution of newly identified compounds across chemical classes for each A–B–O system (A on the x-axis, B on the y-axis) [158]; (b) Base-10 logarithm of the pair-correlation function $g_{ab}$ for each ion pair (A, B) [159].

Knowledge of composition and structure is essential for discovering new materials. Predicting crystalline structures remains a major challenge due to the complexity of combinatorial atomic configurations in three-dimensional space and the high computational demands of first-principles methods [203]. Traditional approaches like structural screening and first-principles calculations require extensive resources, leading to inefficiencies. To enhance structure prediction accuracy, advanced techniques such as random sampling [204-207], metadynamics [208], and minima hopping [209] have been introduced, though they remain computationally expensive [2010]. Consequently, machine learning (ML) methods are increasingly used to explore structural and compositional space more efficiently.

Assessing the thermodynamic stability of new materials is crucial, as stable materials do not decompose into other phases. While energy calculations provide insight, the convex hull method offers a more accurate thermodynamic stability assessment by considering potential decomposition pathways. However, first-principles methods do not account for kinetic effects, which ML can incorporate, making stability predictions more reliable. Schmidt et al. [211] presented a dataset of ~25000 cubic perovskites, excluding noble gases and lanthanides (Figure 9a-c), and applied ML models ridge regression, random forests, extremely randomized trees, and neural networks alongside DFT. Extremely randomized trees provided the best performance, with a mean absolute error (MAE) of 0.121 eV/atom. Prediction accuracy was strongly composition-dependent, especially for first-row elements.

Expanding beyond perovskites, Jha et al. [47, 212] applied ML to ternary compounds in the $AB_2C_2$ space (Figure 9d). Using prototype structures 10-$CeAl_2Ga_2$ and tP10-$FeMo_2B_2$, they identified ~2100 thermodynamically stable systems, 215 of which were absent from existing databases. Most were metallic and non-magnetic, with false-negative rates of 9% (tI10) and 0% (tP10). ML reduced computational costs by ~75%. The research team also introduced ElemNet, a deep neural network predicting minimum formation enthalpy from elemental composition, trained on 275000 compounds from the OQMD database. Compared to conventional ML models using physical descriptors, ElemNet offered superior accuracy (0.050 ± 0.0007 eV/atom), a 9% mean absolute deviation (0.550 eV/atom), and a prediction time of just 0.08 s using GPUs. For 90% of compounds, the formation enthalpy error was below 0.120 eV/atom, demonstrating ElemNet's efficiency, especially for limited datasets.

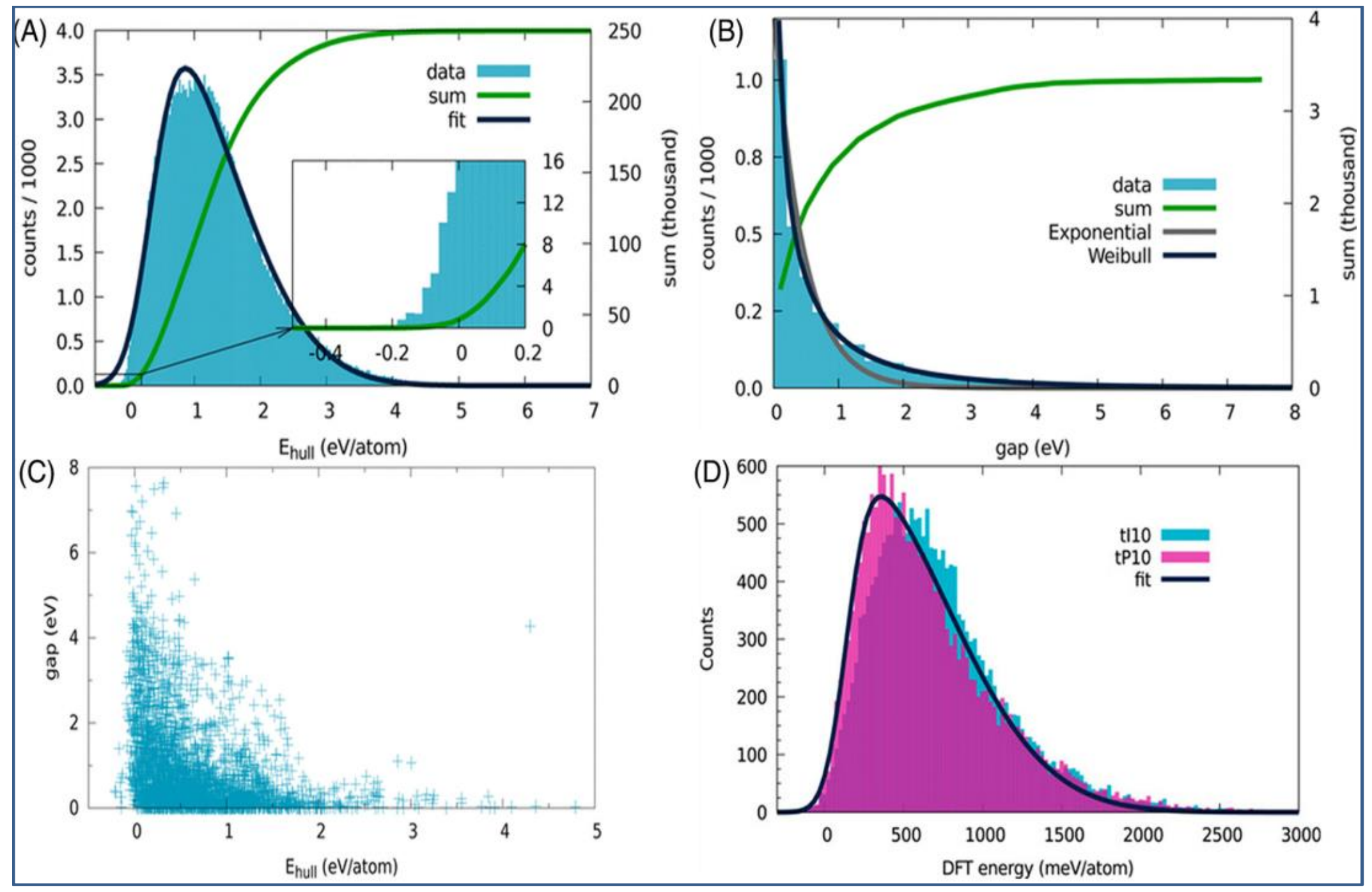

**Figure 9.** Statistical distributions and stability relationships in the cubic perovskite dataset: (a,b) Histograms of $E_{corpus}$ and minimum band gap distributions for ~250000 cubic perovskites (bin size: 25 meV/atom); (c) Scatter plot of band gap as a function of the convex hull energy stability for semiconductor phases [211]; (d) Histogram of $E_{corpus}$ for tI10 training datasets (bin size: 25 meV/atom) [47].

Machine learning has become a powerful tool for predicting the thermodynamic stability of 2D materials and accelerating material discovery. Schleder et al. [213] used ML to classify material stability and predict formation energy for 2,685 non-magnetic materials from the Computational 2D Materials Database (C2DB). By combining SISSO and XGBoost models, they achieved high AUC values for low-, medium-, and high-stability classes (0.93, 0.89, and 0.94, respectively). The most significant features for stability prediction were electron affinity and periodic table group classification. To expand the search space, Meredig's group developed a heuristic ML method for predicting thermodynamic stability in compounds with random compositions [214]. Using DFT data from over 15000 compounds in ICSD, they analyzed 1.6 million potential compositions, identifying 4,500 new stable ternary compounds, 89% of which were novel.

Predicting material composition is a widely used approach in new material discovery, significantly improving the efficiency of screening potential compounds. Zhao et al. [215] applied random forest (RF) and multilayer perceptron neural networks (MLPNN) to predict crystalline systems and space groups based on Magpie, atomic vector, and atomic frequency features. The study showed that RF models performed best in multi-class classification and polymorphism prediction, with the Synthetic Minority Over-sampling Technique (SMOTE) used to address data imbalance. Oleynik et al. [216] used ML to predict Heusler compounds, identifying 12 new stable structures, including $MRu_2Ga$ and $RuM_2Ga$ (M = Ti-Co), with high accuracy (true-positive rate of 0.94). Li et al. [217] optimized medium-entropy alloy (MEA) compositions ($Cr_xCo_yNi_{100-x-y}$) using MD simulations and artificial neural networks (ANN). Their model, trained on 186 datasets and tested on 45, predicted $Co_{21}Cr_{20}Ni_{59}$, $Co_{29}Cr_{30}Ni_{41}$, and $Co_{49}Cr_{30}Ni_{21}$ alloys as low-, medium-, and high-strength compositions with less than 2% error. Chang et al. [218] applied ANN to predict high-entropy alloy (HEA) compositions for maximum hardness, generating five HEA variants, with ML5 ($Al_{24}Co_{18}Cr_{35}Fe_{10}Mn_{7.5}Ni_{5.5}$) showing the highest hardness.

Crystal structure prediction, though computationally intensive, benefits from ML integration. Yang et al. [219] developed a model combining structural prediction with charge mobility analysis to screen 28 isomeric molecules for organic semiconductors, identifying two candidates with high electron mobility. Greiser et al. [220] used the $H_2O$ FLOW ML program to analyze 24,913 compounds from the Pearson Crystal Database (PCD), achieving classification accuracy between 85% and 97%. Hong et al. [221] improved crystal structure prediction by training on unordered structural data. Using neural network potentials (NNP) and first-principles DFT-based methods, they analyzed $Ba_2AgSi_3$, $Mg_2SiO_4$, $LiAlCl_4$, and $InTe_2O_5F$, achieving Pearson correlation coefficients between 0.769 and 0.977 and root mean square errors between 10.7 and 63.7 meV/atom, demonstrating the potential of ML for accurate metastable compound prediction (Figure 10). The NNP method predicted 10-20 low-energy structures within a short computational time. The obtained results confirmed that NNP provides more accurate predictions of low-energy metastable crystalline structures compared to the DFT method.

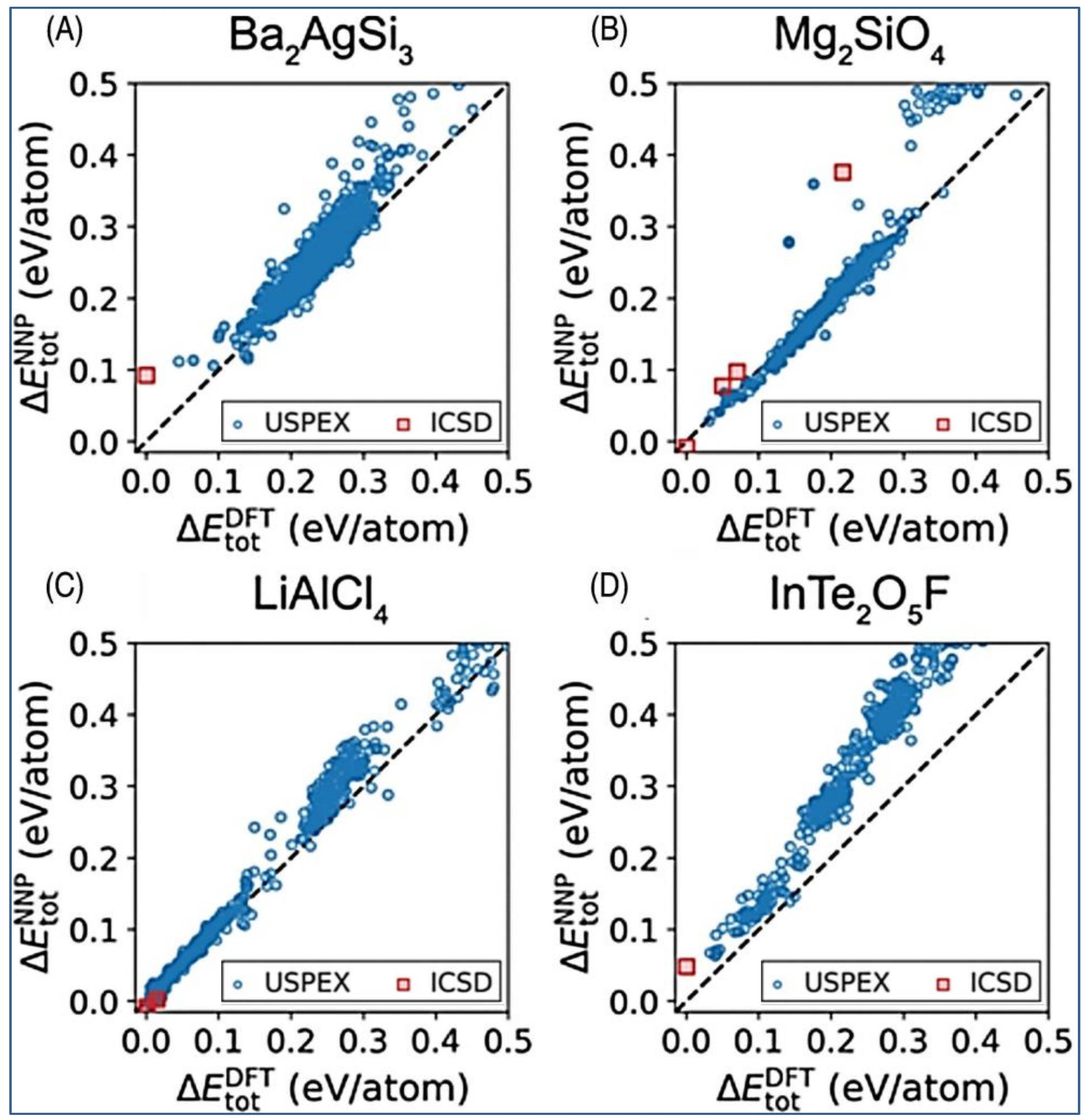

**Figure 10.** Correlations between predicted NNP and calculated DFT energies for the: (a) $Ba_2AgSi_3$; (b) $Mg_2SiO_4$; (c) $LiAlCl_4$ and (d) $InTe_2O_5F$; blue circles represent metastable structures from USPEX, while red squares represent experimental structures from ICSD [221].

Atomic catalysts (ACs) have become a major focus in electrochemistry due to their high electroactivity and broad applications in energy storage and conversion. Traditional trial-and-error approaches for selecting catalyst-substrate combinations are inefficient, making machine learning (ML) a powerful alternative [222]. Huang et al. applied ML to study graphdiyne-based (GDY) electrocatalysts, using the redox barrier model to evaluate electron transfer efficiency. DFT calculations identified the most promising neutral atom-based electrocatalysts, which were experimentally validated (Figure 11a) [223]. They expanded this method to hydrogen evolution reactions (HER), using a tree-based ensemble learning approach (bagging trees) to predict key adsorbate binding energies, showing a strong correlation with DFT results (Figure 11b, c) [224]. Further, Huang's team investigated bimetallic GDY electrocatalysts (GDY-DAC), screening 990 metal combinations with a geostatistical radial basis function (georadar) algorithm. Their model accurately predicted formation energies with a low RMSE of 0.16 but faced challenges in d-band center predictions (Figure 11d-f) [225]. Expanding the study to s- and p-orbital metals [226], they found that these elements reduced predictive accuracy, particularly for alkali and alkaline earth metals, which exhibited significantly higher RMSE values (Figure 11g-i). These findings emphasize ML's role in catalyst discovery while highlighting challenges in modeling electronic structures.

Li et al. [227] proposed an adaptive machine learning method to accelerate the discovery of cubic perovskite electrocatalysts ($ABO_3$) for the oxygen evolution reaction (OER). Using Gaussian process methods with electronic structure and composition as input parameters, the model efficiently analyzed ~4000 $AA'B_2O_6$-type perovskites, identifying ten previously unstudied stable compounds with high catalytic activity. The method also minimized prediction uncertainty with low computational costs. $CO_2$, a renewable feedstock for fuel synthesis, has been explored using ML to identify efficient electrocatalysts. Malek et al. [228] applied classification and regression algorithms, including random forests, to optimize $CO_2$ reduction processes. The random forest model provided the most accurate predictions, leading to the identification of Pt and nickel-stabilized yttria-zirconia dioxide (Ni-YSZ) as optimal materials for low- and high-temperature $CO_2$ reduction.

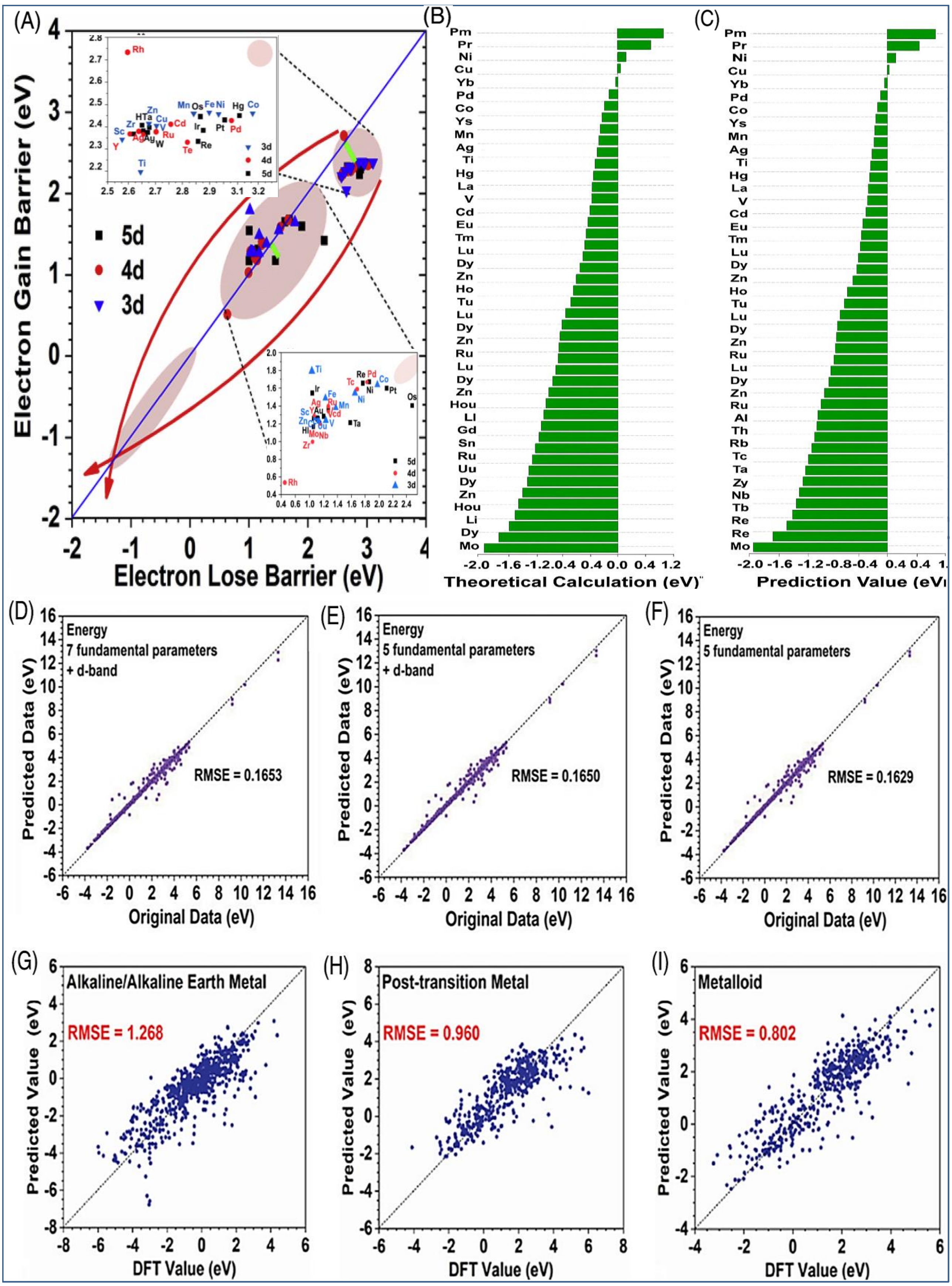

**Figure 11.** Electron transfer and formation-energy modeling in TM–GDY and GDY-DAC systems: (a) Electron-transfer evolution in TM–GDY [223]; (b,c) comparison of theoretical vs. ML-predicted 2-h adsorption values [224]; (d–f) formation-energy predictions using different intrinsic parameters

and the d-band center [225]; (g–i) Georadar-model predictions vs. original formation-energy data for alkali/alkaline-earth metals, post-transition metals, and metalloids in GDY-DAC [226].

ML has significantly accelerated atomic catalyst research by reducing experimental workload, making it a key tool for discovering efficient electrocatalysts. It enables predictions of composition, crystalline structure, and thermodynamic stability, proving invaluable for materials science. However, limitations remain, particularly the scarcity of high-quality experimental data due to costly and extreme-condition research. Addressing this challenge requires increased funding, adherence to FAIR data-sharing principles, and high-precision theoretical modeling to supplement existing datasets. While ML expedites material discovery, overcoming data limitations is essential for broader applications.

## 6. ML-driven autonomous experiments in materials science

Automated laboratories equipped with artificial intelligence (AI) and robotic systems are transforming modern chemistry and materials science by conducting experiments, analyzing data, and optimizing processes with minimal human intervention. These intelligent "Robot scientists" accelerate the discovery of novel materials, optimize synthesis conditions, and enhance high-throughput screening capabilities. Studies [229-231] have demonstrated ML-driven robotic platforms optimizing chemical reactions, with the study [229] autonomously performing over 1000 experiments to refine a cross-coupling reaction. Shields, B. et al. [230] utilized AI to predict reaction yields, accelerating the optimization process tenfold, while Granda, J.et al. [231] identified new catalytic systems through real-time adaptive experimentation. Automated material discovery has also benefited from ML integration. Studies [232-234] applied robotic systems to synthesize and test new catalysts and functional materials, Li, Z. et al. [232] conducting 5000 autonomous experiments for hydrogen energy applications, leading to the identification of a record-breaking catalyst. Tabor, D. et al. [233] leveraged AI for data-driven material property predictions, reducing experimental requirements by 70%.

High-throughput screening, essential for materials science and drug discovery, has been significantly enhanced by ML-driven automation. Studies [235-237] employed AI-guided platforms to evaluate thousands of compounds, with Hensenet al. [235] screening 10000 materials in one week and discovering a potential pharmaceutical candidate. Study [161] integrated ML analyzes screening data, accelerating the identification of active compounds. Studies [238-240] applied AI-enhanced robotic systems for organic synthesis, with Coley et al. [238] automating the production of over 100 compounds in a few days a task that would traditionally take months. Segler, M. et al. [239] optimized reaction parameters, increasing product yield by 30%, while a study [240] automated complex molecule synthesis and purification.

Beyond synthesis and screening, ML plays a pivotal role in autonomous experiment management and real-time data analysis. Studies [241-243] demonstrated AI-powered systems capable of extracting insights from vast datasets and guiding experimental decisions. Kusne, A. et al. [241] applied AI to analyze X-ray diffraction data and predict material structures, while Stein, H. et al. [242] employed reinforcement learning to dynamically adjust experiment conditions. ML-enhanced robotic platforms have also accelerated energy storage research. Studies [244-246] explored novel battery materials, with Dave, A. et al. [244] conducting 2000 experiments to identify an improved electrode material. Sendek, A. et al. [245] used predictive models to optimize electrolyte properties, cutting development time by 50%. Similarly, studies [247-249] automated composite material synthesis, with study [247] evaluating 1000 compositions to identify a material with superior mechanical properties, while study [248] reduced experimental complexity by 60% through AI-guided decision-making.

Despite these advances, scaling autonomous ML-driven laboratories remains a challenge due to the need for adaptable and modular platforms. In 2021, Leroy Cronin’s group [250] introduced the Chemputer, an ML-integrated system capable of executing multiple reaction types, including solid-phase peptide synthesis (SPPS) and iterative cross-coupling (ICC). Their platform executed ~8500 operations across 10 modules, reusing only 22 steps, demonstrating a universal framework for reaction automation. Compared to conventional automation systems, the key innovation lies in its ability to conduct multiple sequential reactions autonomously, improving experimental reproducibility and standardization. However, optimizing reaction conditions remains labor-intensive, as current AI-driven systems are most effective for predefined synthesis protocols rather than exploratory research [251]. The integration of self-learning AI models capable of hypothesis generation and iterative refinement will be essential for the next generation of autonomous laboratories. Future advancements in ML-driven experimental automation will transform materials science by accelerating discoveries, reducing costs, and facilitating the development of innovative materials with unprecedented properties.

## 7. ML and acceleration of interdisciplinary research in materials chemistry

With the advancement of ML methods, chemistry, and materials science is entering a new era where high-performance computing, virtual screening, and automated laboratories accelerate the discovery of novel compounds and materials. ML not only assists in selecting compounds for synthesis but also predicts new experiments based on collected data. The 15$^{th}$ ASLLA Symposium on “Accelerated Chemistry with AI” (September 25-28, 2022, KIST, South Korea) brought together 45 researchers to discuss key topics such as data, new applications, ML algorithms, and education, which form the foundation of this review. The quality and scale of data play a critical role in developing robust ML models. Constructing extensive databases requires evaluating data diversity and novelty [252]. ML efficiency improves through multi-fidelity learning [253], uncertainty quantification, delta learning, and active learning, which enhance predictive capabilities even with limited data [254,255]. Integrating physical rules into ML models and using entropy-based sampling help analyze the vast chemical space more effectively [256,257]. However, data accessibility remains a challenge. For example, databases like QM9 require additional calculations to obtain key properties, making them difficult for ML experts without a strong chemistry background. Developing web-based interfaces and standardized formats can facilitate usability [258]. In this regard, dynamic community databases, modeled on the Common Task Framework (CTF) used in bioinformatics, could foster interdisciplinary collaboration [259-261].

Large chemical reaction databases such as USPTO, Reaxys, and SciFinder contain extensive datasets, but their diversity, quality, and accessibility are limited. Efforts to create an Open Reaction Database face challenges due to the lack of high-quality data [262-265]. To improve ML efficiency in organic synthesis, delta learning, transfer learning, and few-shot learning are essential [266,267]. However, the absence of failed experiment data restricts the ability of ML models to generate novel reaction pathways. A cultural shift toward documenting and publishing all experimental results both successful and unsuccessful could address this issue [268]. ML is also widely used in modeling non-equilibrium states, allowing the exploration of data beyond known stable structures and predicting reaction mechanisms [269,270]. Despite advancements in computational methods for materials science, predicting synthesizability remains a challenge. To bridge the gap between computational discoveries and experimental validation, integrated autonomous workflows that combine ML-driven predictions with real-world experiments are necessary [271]. The bias toward publishing only successful experiments results in ML models being trained on a limited set of structures, restricting their ability to discover truly novel compounds [272]. Incorporating metadata from unsuccessful experiments could significantly improve predictive accuracy.

Beyond optimization, ML fosters creativity in chemistry, enabling the discovery of novel molecular structures and unconventional reaction pathways. The rise of large language models (LLMs) raises fundamental questions about whether ML can truly “understand” scientific concepts based on data alone [273,274]. ML is also solving challenges in multi-scale material design, optimizing properties at both the atomic level (catalysis, thermoelectrics) and the macroscopic level (mechanical strength, operational stability) [275-278].

The development of self-driving laboratories allows for real-time adaptation of experiments, significantly reducing costs. However, building these facilities requires substantial investment. As an alternative, virtual laboratories that combine ML-driven simulations with physical experiments are being explored [279]. To enhance predictive accuracy in chemistry, ML algorithms must be specifically designed for chemical data. For example, physically informed architectures are already being used for molecular property prediction [253]. However, most ML models have an interpolative nature, limiting their ability to predict materials beyond known datasets [280]. Advancing materials science requires multi-objective optimization algorithms that account for multiple criteria in discovering new compounds [280-282].

ML should not only accelerate the discovery of new compounds but also drive fundamental breakthroughs in chemistry. For instance, recent studies on autonomous ML-driven systems have demonstrated the potential for serendipitous discoveries, leading to novel chemical reactions [283]. At the same time, improving uncertainty quantification is essential, as active learning relies on the precise calibration of epistemic and aleatoric uncertainty [256,284]. The expansion of ML in chemistry requires a fundamental shift in education. Courses on data science, machine learning, and computational chemistry should be integrated into university curricula [285,286]. Hands-on training in statistical analysis, FAIR data management, and ML-driven automation is essential [287,289]. Virtual reality (VR) and remote laboratories could further enhance student engagement [290,291]. One of the biggest challenges is the lack of faculty training and the conservative nature of academic curricula. Proposed solutions include summer coding boot camps, automation-focused courses, and incorporating ML techniques into laboratory training [292]. Meanwhile, chemistry faces a declining interest among students, as many are drawn to computer science fields. However, the development of autonomous laboratories and ML-driven tools could enhance the appeal of chemistry by highlighting its contributions to green chemistry, pharmaceuticals, and space exploration [293,294].

In conclusion, machine learning is transforming chemistry and materials science, accelerating research, reducing costs, and opening new interdisciplinary avenues. The future lies in dynamic databases, standardized datasets, and self-driving laboratories, which will enable not just quantitative advancements but also conceptual breakthroughs in scientific discovery.

Next-generation functional materials extend beyond semiconductors and batteries to include biologically inspired systems. For example, ML has been applied to design self-healing polymers, artificial muscles, and bio-inspired photonic structures. These materials draw from biological processes to achieve multifunctionality and adaptability [17, 24, 36, 134, 275].

## 8. Challenges and prospects

The fourth paradigm of science, driven by data, has significantly transformed materials science research, enabling the prediction of material properties, the discovery of novel compounds, and the design of functional materials. Machine learning (ML) has introduced powerful tools that surpass traditional experimental and computational methods in flexibility, accuracy, and generalization capabilities. However, despite these advancements, ML in materials science still faces several challenges and limitations [125, 295-305] that require further investigation (Figure 12).

One of the primary challenges is data scarcity and uneven distribution. Existing databases, such as Materials Project, OQMD, AFLOW, and NOMAD, predominantly focus on inorganic crystalline materials, while amorphous materials, polymers, composites, and defect-rich structures remain underrepresented. This lack of data limits ML models' ability to predict the properties of these materials. Additionally, datasets often contain noise, missing values, and inconsistencies, making preprocessing and standardization essential [295]. The variability of experimental conditions further complicates the alignment between computational predictions and real-world measurements. To address this, synthetic data generation techniques using generative adversarial networks and active learning are being actively developed to supplement datasets. Active learning selects the next most informative experiments or calculations in a closed loop, focusing data collection where model uncertainty is high and expected improvement is maximal; in materials discovery, adaptive design has accelerated the search for targeted properties and has even been deployed in autonomous labs that realized dozens of previously unrealized compounds within days [48,49]. Generative adversarial networks, in turn, learn the data distribution to produce statistically realistic synthetic samples (structures, microstructures, or spectra), which can balance under-represented classes and enlarge training sets; for example, GAN-based augmentation improved hardness prediction in high-entropy alloys and enabled statistically equivalent alloy microstructure generation [306,307]. In modern workflows these approaches are complementary: a generative model proposes candidates, while an active-learning policy prioritizes which candidates (or conditions) to evaluate next, thereby reducing the number of DFT calculations/experiments and shortening time-to-target [308,309].

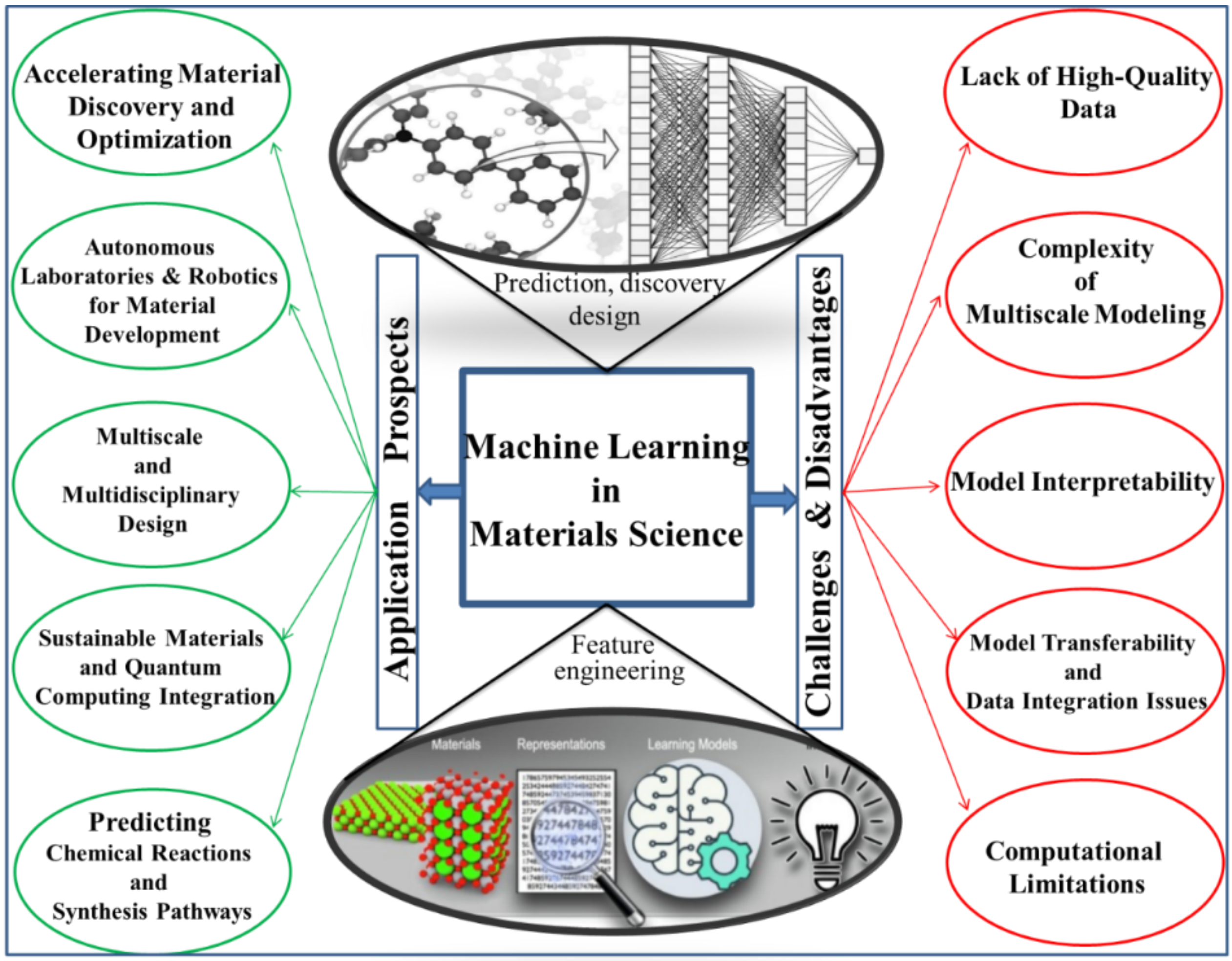

**Figure 12.** Schematic illustration of the main challenges and prospects for the application of machine learning in materials science.

A recent survey provides a consolidated view of generative AI methods and practical guidance for their safe use in inverse design [21]. Together with ongoing efforts to curate standardized, merged databases, these strategies directly mitigate data scarcity and sampling bias. In parallel, data quality and quantity governance frameworks for materials ML balance sample size with descriptor dimensionality under domain constraints, improving downstream robustness [311,312]. For text-mining pipelines, standardized construction of high-quality corpora substantially improves extraction accuracy and reproducibility [313]. Establishing standardized databases that integrate both experimental and theoretical high-throughput computational data is a promising approach [296,297,309].

Another major issue is the low interpretability of ML models, particularly deep neural networks, which often function as “black boxes”. Since ML predictions stem from complex mathematical operations, they must be validated through experiments, positioning ML as a hypothesis-generation tool rather than a standalone solution [3,5]. Methods from explainable artificial intelligence (XAI) are being developed to improve interpretability. SHAP (Shapley Additive Explanations) evaluates the contribution of individual features (chemical composition, crystal structure) to ML predictions [4]. LIME (Local Interpretable Model-Agnostic Explanations) allows localized analysis of model behavior [6]. Integrated gradients and physics-informed neural networks (PINNs) enhance model reliability by incorporating established physical laws, which is particularly valuable in studying superconductors and catalysts [51,142]. Additionally, the development of interpretable descriptors with physical significance, such as the one-dimensional tolerance factor ($\tau$) proposed by Bartel et al. for perovskite stability prediction (92% accuracy), bridges the gap between ML predictions and fundamental materials understanding [298, 299]. For solid state electrolytes hierarchically encoding crystal structure HECS with causal inference integrates global and local Li+ conduction environments to identify key descriptors, enabling accurate activation energy prediction in argyrodites and transferable rational design rules [314, 315]. Despite these efforts, XAI methods can provide conflicting interpretations, and their high computational cost slows large-scale data analysis [83]. However, integrating XAI with autonomous laboratories enhances predictability and efficiency in materials discovery [243].

An emerging approach to improving ML efficiency in materials science is automated model selection and tuning via AutoML, which optimizes algorithm selection, hyperparameters, and data preprocessing. AutoML significantly reduces the time and computational resources needed for model development, making it particularly useful in complex materials science problems. For example, AutoML can identify the most suitable models for predicting thermoelectric materials or catalysts, ensuring better generalization when experimental data are unavailable [300-303]. This approach is especially beneficial in low-data environments and for tasks requiring rapid model adaptation [77-95,304].

The integration of quantum computing with ML presents another promising avenue for materials science. Quantum machine learning (QML) algorithms have the potential to accelerate model training and quantum system simulations, particularly for electronic structure calculations. Quantum algorithms can optimize exchange-correlation functionals in density functional theory (DFT), enhancing computational accuracy and efficiency beyond classical methods [305].

Despite these challenges, ML is continuously transforming materials science, accelerating the discovery, optimization, and design of advanced materials [315-324]. Addressing key issues such as data limitations, model interpretability, and computational constraints will unlock the full potential of ML. The development of standardized databases, interpretable descriptors, and hybrid ML-traditional approaches is paving the way for breakthroughs in functional materials, nanotechnology, and bioengineering. In the future, ML may not only revolutionize materials discovery but also play a crucial role in uncovering fundamental physical principles that govern material behavior, marking a paradigm shift in materials science.

## 9. Conclusions

Machine learning (ML) has emerged as a transformative paradigm in modern materials science, dramatically accelerating the prediction, design, and discovery of next-generation materials. Over the past decade, remarkable progress has been achieved through the application of ML algorithms to analyze large and diverse datasets, including structural, spectral, and experimental inputs, enabling faster and more accurate modeling of complex material behaviors. Advanced algorithms such as gradient boosting, deep neural networks, and graph neural networks (GNNs) have notably improved the precision of property predictions, while generative models such as GANs and VAEs have begun to demonstrate their value in the inverse design of novel compounds.

ML-driven approaches have contributed to breakthroughs across multiple domains, including the development of metamaterials with unconventional optical responses, the discovery of high-performance thermoelectric and superconducting materials, and the optimization of electrode materials for energy storage applications. Furthermore, the integration of ML with quantum-mechanical simulations (e.g., DFT), high-throughput computations, and experimental automation has substantially reduced the time and cost associated with materials innovation.

Looking forward, the role of ML in materials science is expected to deepen. The rise of AutoML frameworks, robotic laboratories, and self-driving experimentation platforms promises to enable closed-loop material design cycles with minimal human intervention. Quantum machine learning (QML) offers the potential to simulate strongly correlated systems and topologically complex materials with unprecedented fidelity. Meanwhile, interpretable ML models and explainable AI (XAI) approaches such as SHAP and LIME are essential for fostering trust, transparency, and scientific insight, especially in safety-critical and industrial applications.

In addition, the emergence of FAIR-compliant datasets (Findable, Accessible, Interoperable, Reusable) and efforts to curate diverse and balanced materials databases will be vital for ensuring reproducibility and model generalization. Expanding the reach of ML to biologically inspired, hybrid organic–inorganic, and soft-matter systems represents another frontier, where traditional simulation approaches remain limited.

Ultimately, combining machine learning with domain knowledge and modern computational and experimental platforms is improving screening efficiency, prediction fidelity, and experiment planning in materials research. As these tools mature, we expect steady, measurable gains in designing materials with target properties and in translating candidates to prospective validation. Priorities are standardized multi-fidelity datasets that include polymers and defect-rich systems, calibrated uncertainty and interpretability, and closed-loop active learning with reproducible benchmarks. Progress on these fronts will support dependable use in applications across energy, sustainability, and health.

## ABBREVIATIONS

**AFLOW** – Automatic Flow for Materials Discovery
**ANN** – Artificial Neural Network
**AutoML** – Automated Machine Learning
**BMG** – Bulk Metallic Glass
**CatBoost** – Categorical Boosting
**CGCNN** – Crystal Graph Convolutional Neural Network
**CNN** – Convolutional Neural Network
**DBSCAN** – Density-Based Spatial Clustering of Applications with Noise
**DFT** – Density Functional Theory
**DCSA** – Data-Driven Classification of Superalloys
**DT** – Decision Tree
**FDTD** – Finite-Difference Time-Domain
**GAN** – Generative Adversarial Network
**GNN** – Graph Neural Network
**HEA** – High-Entropy Alloy
**H2O AutoML** – H2O Automated Machine Learning Platform
**HOIP** – Hybrid Organic-Inorganic Perovskite
**ICSD** – Inorganic Crystal Structure Database
**IR** – Infrared Spectroscopy
**KNN** – K-Nearest Neighbors
**LIME** – Local Interpretable Model-agnostic Explanations
**LSTM** – Long Short-Term Memory
**MAE** – Mean Absolute Error
**MD** – Molecular Dynamics
**MIPHA** – Microstructure and Property Prediction of High-Entropy Alloys
**ML** – Machine Learning
**MOF** – Metal-Organic Framework
**OQMD** – Open Quantum Materials Database
**OLED** – Organic Light-Emitting Diode
**OPTICS** – Ordering Points To Identify the Clustering Structure
**PCA** – Principal Component Analysis
**PINN** – Physics-Informed Neural Network
**PLMF** – Property-Labeled Materials Fragment
**PUK** – Pearson VII Universal Kernel
**QMD** – Quantum Molecular Dynamics
**$R^2$** – Coefficient of Determination
**RF** – Random Forest
**RNN** – Recurrent Neural Network
**RMSE** – Root Mean Square Error
**ROC** – Receiver Operating Characteristic
**SHAP** – Shapley Additive Explanations
**SVM** – Support Vector Machine

**t-SNE** – t-distributed Stochastic Neighbor Embedding
**TPOT** – Tree-based Pipeline Optimization Tool
**UMAP** – Uniform Manifold Approximation and Projection
**VAE** – Variational Autoencoder
**XAI** – Explainable AI
**XGBoost** – eXtreme Gradient Boosting
**XRD** – X-ray Diffraction

## Acknowledgements

This work was supported by the International Science and Technology Center (grant no. TJ-2726). The authors express their deep gratitude to the Government of Japan for the financial support of this project and the creation of a modern laboratory at the S.U. Umarov Physical-Technical Institute of the National Academy of Sciences of Tajikistan